\documentclass[12pt]{JHEP3}
\usepackage{amsmath}

\newcommand{\be}{ \begin{equation}}
\newcommand{\ee}{\end{equation}}
\newcommand{\bea}[1]{\begin{eqnarray}\label{#1} }
\newcommand{\eea}{\end{eqnarray}}

\def\ZZZ{{\hskip-3pt\hbox{ Z\kern-1.6mm Z}}}
\def\zzz{{\hskip-3pt\hbox{ z\kern-1mm z}}}

\def\one{{\hbox{ 1\kern-.8mm l}}}
\def\zero{{\hbox{ 0\kern-1.5mm 0}}}

\title{Large ${\cal N}=4$ Holography}

\author{
Matthias R.\ Gaberdiel$^{a}$ and Rajesh Gopakumar$^{b}$ \\ 
$^a$Institut f\"ur Theoretische Physik, ETH Zurich, \\
$\;$CH-8093 Z\"urich, Switzerland \\
$\;$\email{gaberdiel@itp.phys.ethz.ch}\\ \\ 
$^b$Harish-Chandra Research Institute, \\
$\;$Chhatnag Road, Jhusi,\\
$\;$Allahabad, India 211019\\
$\;$\email{gopakumr@hri.res.in}}

\abstract{The class of 2d minimal model CFTs with higher spin AdS$_3$ duals is extended to  theories with large ${\cal N}=4$ superconformal symmetry. We construct a higher spin theory based on the global $D(2,1|\alpha)$ superalgebra, and propose a large $N$ family of cosets as a dual CFT description. We also indicate how a non-abelian version of this Vasiliev higher spin theory might give an alternative description of IIB string theory on an ${\rm AdS}_3\times {\rm S}^3\times {\rm S}^3\times {\rm S}^1$ background.}

\begin{document}

\section{Introduction and Summary}

Our understanding of string theory provides the clearest rationale behind the existence of gauge-gravity (or more generally, gauge-string) dualities. The dual descriptions of D-branes, as demanded by the internal consistency of string theory, lies behind all the best understood examples of the AdS/CFT correspondence. However, starting with the work of Klebanov-Polyakov and others, a new class of AdS/CFT dualities \cite{Klebanov:2002ja, Sezgin:2003pt, Giombi:2009wh, Giombi:2010vg, Gaberdiel:2010pz, Aharony:2011jz,Giombi:2011kc} have been uncovered which do not obviously arise from any embedding in string theory. This is related to some of their special features. In the first place, they are often genuinely non-supersymmetric (with only bosonic degrees of freedom).  Secondly, the AdS bulk description is not in terms of string theory (or supergravity), but rather in terms of  a Vasiliev system of equations \cite{Vasiliev:2003ev} for a tower of massless higher spin gauge fields, thus having far fewer degrees of freedom than a string theory. This is also reflected in the third feature that the CFT has only vector like physical degrees of freedom --- any gauge or adjoint degrees of freedom are non-propagating. Fourthly, when the dual field theories are studied on a non-simply connected space (such as ${\rm S}^1$ for the 2d CFTs and a Riemann surface of genus $>1$ for the 3d CFTs) there is an effective continuum of light states that appears in the large $N$ limit
\cite{Gaberdiel:2010pz, Gaberdiel:2011zw, Banerjee:2012gh}. Finally, while the Vasiliev theory provides a successful classical 
description in the bulk, it is not clear, even in principle, how to quantise it. In other words, we do not have any way to 
systematically compute $\frac{1}{N}$ corrections in the bulk. 

While these novel features make such examples fascinating objects of study in themselves, 
it would nevertheless be desirable to know to what extent they can be embedded into string theory. In fact, one of the initial motivations to study the Vasiliev system of equations (in the AdS/CFT context) was the expectation that these might govern a sector of the tensionless limit of string theory on AdS \cite{Sundborg:2000wp, Witten}. More generally, the role of higher spin gauge symmetries (in broken or unbroken form) in string theory is also yet to be elucidated. At the same time, such an embedding will presumably shed light on puzzling aspects of the higher spin/CFT correspondence, for example the last two points of the preceding paragraph.

To make progress in this direction, it is natural to start with a higher spin example with a large amount of supersymmetry, and look for its embedding into string theory. 
Indeed, in the case of the AdS$_4$/Chern-Simons vector model dualities \cite{Aharony:2011jz,Giombi:2011kc} (which generalise the ${\rm O}(N)$ vector model duality \cite{Klebanov:2002ja, Sezgin:2003pt, Giombi:2009wh, Giombi:2010vg}), such a candidate embedding has been proposed \cite{Chang:2012kt}. This relates the ${\cal N}=6$ 
${\rm U}(M)\times {\rm U}(N)$ ABJ theory with a Vasiliev theory having additional ${\rm U}(M)$ Chan-Paton indices. Since the former theory is also believed to possess a string dual, this proposal implies that the string states are built from confined bound states of the non-abelian Vasiliev theory. The simpler vector-like dualities are recovered in the limit when $M=1$ (or more generally when
$M$ is finite), while $N$ is taken to be large.  

In this paper we will take a first step in a similar direction for the case of the AdS$_3$ duals to two dimensional coset CFTs
\cite{Gaberdiel:2010pz} (see \cite{Gaberdiel:2012uj} for a overview). We will identify a higher spin/CFT$_2$ example which, we feel, holds most promise in being embeddable into string theory. As mentioned earlier, it is advantageous to consider highly supersymmetric examples. There are several ${\rm AdS}_3$ string backgrounds with ${\cal N}=2$ supersymmetry. We will therefore consider an example whose symmetry contains the 
so-called large ${\cal N}=4$\footnote{Thus creating confusion and distress amongst those who believe that large $N=3$.} superconformal algebra
 \cite{Sevrin:1988ew, Schoutens:1988ig, Spindel:1988sr, VanProeyen:1989me, Sevrin:1989ce}  in both the bulk and the boundary. There is, essentially, one string background 
with this supersymmetry algebra which has the geometry  ${\rm AdS}_3\times {\rm S}^3\times {\rm S}^3\times {\rm S}^1$. On
the AdS side, higher spin theories with extended supersymmetry were recently also considered in \cite{Henneaux:2012ny}.
 
 The large ${\cal N}=4$ symmetry has four supercharges\footnote{We will be considering only parity invariant theories, and thus our statements should be viewed as holding separately for left and right moving sectors. Thus we have four complex supercharges, both on the left as well as on the right.}, and {\it two} $\mathfrak{su}(2)$ affine algebras. 
This is to be contrasted to the small or regular ${\cal N}=4$ superconformal algebra which contains only a single 
$\mathfrak{su}(2)$ affine algebra. The presence of the two $\mathfrak{su}(2)$ algebras with their individual levels 
$k^{\pm}$ introduces an additional parameter that characterises the large ${\cal N}=4$ algebra --- namely, 
$\gamma =\frac{k^-}{k^++k^-}$. The corresponding superconformal algebra (see appendix~\ref{app:alg} for the detailed form) is 
customarily denoted as $A_{\gamma}$ in the literature. Strictly speaking, the algebra has a linear as well as a non-linear version  (denoted by $\tilde{A}_{\gamma}$), as we will discuss later. In either case, the central charge is constrained to take a specific form
in terms of the levels $k^{\pm}$; for the linear $A_{\gamma}$ algebra it is 
\begin{equation}\label{linagam}
c = \frac{6 \, k^+ k^-}{k^+ + k^-} \ ,
\end{equation}
while for $\tilde{A}_{\gamma}$ we have instead 
$c = \tfrac{6 k^+ k^- + 3 (k^+ + k^-)}{k^+ + k^-+2}$.
\smallskip

We will consider a family of coset CFTs with large ${\cal N}=4$ superconformal symmetry. These take the form
\be\label{N4coset2a}
\frac{\mathfrak{su}(N+2)^{(1)}_{\kappa} }{\mathfrak{su}(N)^{(1)}_{\kappa} \oplus \mathfrak{u}(1)}
\oplus \mathfrak{u}(1) \cong \frac{\mathfrak{su}(N+2)_{k} 
\oplus \mathfrak{so}(4N+4)_1}{\mathfrak{su}(N)_{k+2}\oplus  \mathfrak{u}(1)} \oplus \mathfrak{u}(1) \ ,
\ee
where the left-hand-side refers to a manifestly ${\cal N}=1$ form of the coset for which  $\kappa=k+N+2$. 
As written these cosets contain the linear $A_\gamma$ algebra, and the levels
of the two $\mathfrak{su}(2)$ factors 
are $k^+=(k+1)$ and $k^-=(N+1)$, respectively. For the relation to the higher spin theory, we will 
actually quotient out four of the free fermions of the 
$\mathfrak{so}(4N+4)_1$ algebra together with a $\mathfrak{u}(1)$ factor, leading to the non-linear $\tilde{A}_{\gamma}$ 
algebra \cite{Goddard:1988wv}. 
The full coset algebra then forms an 
extended algebra of higher spin conserved currents, forming a ${\cal W}$-algebra that contains $\tilde{A}_\gamma$
as a subalgebra; we shall denote this resulting ${\cal W}$-algebra as 
$s\tilde{\cal W}^{(4)}_{\infty}[\gamma]$.

As in other examples of minimal model holography, we will take a large $N$, $k$ limit of the coset keeping the ratio 
\be\label{costhft}
\lambda =\frac{N+1}{N+k+2} = \gamma 
\ee
fixed. This limit appears to be sensible just like in other large $N$ cosets with which it shares most qualitative features. This includes a set of primaries which are candidate single particle states together with a near continuum of light states. 
\medskip

The `wedge' or global part of the large ${\cal N}=4$ algebra $\tilde{A}_\gamma$ is 
a finite dimensional Lie superalgebra  known as $D(2,1|\alpha)$, 
where the parameter $\alpha=\frac{\gamma}{1-\gamma}$ 
(see Appendix~\ref{app:D21} for
a brief introduction to $D(2,1|\alpha)$). Furthermore, the 
wedge part of the full ${\cal W}$-algebra $s\tilde{\cal W}^{(4)}_{\infty}[\gamma]$ is a higher spin extension of 
$D(2,1|\alpha)$. It is to be identified with the global symmetry of a higher spin theory on AdS$_3$; in fact,
the corresponding higher spin theory can simply be formulated as a Chern-Simons theory based on this wedge algebra.
We shall  argue that this higher spin extension of $D(2,1|\alpha)$ can be identified with an explicit algebra 
shs$_2[\mu]$ that we construct (where $\mu=\frac{\alpha}{1+\alpha}$), see Section~\ref{sec:hspin}.
In particular, we shall show that the symmetries, as well as the gross features of the spectrum match those of the
large $N$ 't~Hooft limit of  the coset (\ref{N4coset2a}).
The parameters on both sides are simply related via $\mu=\gamma=\lambda$. This therefore adds to the growing list \cite{Gaberdiel:2010pz, Ahn:2011pv, Gaberdiel:2011nt, Creutzig:2011fe, Creutzig:2012ar,Beccaria:2013wqa} of higher spin AdS$_3$/CFT$_2$ dualities.
\smallskip

String theory with large ${\cal N}=4$ superconformal symmetry 
\cite{Boonstra:1998yu,Elitzur:1998mm, de Boer:1999rh, Gukov:2004ym} has been 
somewhat of an outlier in the AdS/CFT correspondence. As mentioned, there is only one known background of superstring theory, with the geometry
AdS$_3\times {\rm S}^3\times {\rm S}^3\times {\rm S}^1$, which possesses the large ${\cal N}=4$ superconformal symmetry. 
However, there is no complete proposal for a dual CFT, and even the partial proposals have problems as explained in \cite{Gukov:2004ym}. This leaves open the appealing possibility of using a non-abelian extension of the shs$_2[\mu]$ Vasiliev theory and a corresponding generalisation of the coset  (\ref{N4coset2a}) to provide an alternative description of this string theory background. We will see some encouraging signs that this might indeed be the case. 

In particular, we shall find 
that a non-abelian version of the Vasiliev theory which we have constructed, has a BPS spectrum which matches 
with that of the string theory --- even this was difficult to see in the extant proposals for the string theory based on a symmetric product of ${\rm S}^3\times {\rm S}^1$  \cite{Gukov:2004ym}. Note that the Vasiliev theory has a large extended unbroken ${\cal W}$-symmetry
and hence can potentially describe the string theory only at a special point in its moduli space. The Vasiliev theory does possess a corresponding marginal deformation which preserves the large ${\cal N}=4$ superconformal symmetry, 
but is likely to break the higher spin symmetry.   
We do not, however, yet have a concrete proposal for a coset generalising (\ref{N4coset2a}), which would be a candidate dual to the non-abelian Vasiliev theory. The constraints of preserving the  large 
${\cal N}=4$ superconformal algebra imposes strong constraints on any candidate, and at least some of the 
obvious generalisations seem to be unsatisfactory. It would be very interesting to find a suitable coset that satisfies
all of these requirements.
\medskip

The paper is organised as follows. We will describe the construction of the higher spin algebra  shs$_2[\mu]$ based on 
$D(2, 1| \alpha)$ and the resulting Vasiliev theory, in Section~\ref{sec:hspin}. Next we discuss, in 
Section~\ref{sec:coset}, the coset (\ref{N4coset2a}) and some of its properties including the realisation of the large 
${\cal N}=4$ superconformal symmetry. 
Section~\ref{sec:spec} describes the spectrum of primaries of the coset concentrating on the BPS states. This brings us to a 
position (see Section~\ref{sec: compar})
where we can compare the states in the Vasiliev theory with the large $N$, $k$ limit of the coset spectrum of 
Section~\ref{sec:spec}. In Section~\ref{sec:stringy} we describe why a non-abelian version of the Vasiliev theory of 
Section~\ref{sec:hspin} can potentially be equivalent to string theory on 
AdS$_3 \times {\rm S}^3\times {\rm S}^3\times {\rm S}^1$ at a special point in its  moduli space. We also 
outline some of the constraints on coset generalisations. Finally we conclude in Section~\ref{sec:concl}. The technical 
appendices describe details of $D(2,1|\alpha)$, the large ${\cal N}=4$ algebra and its non-linear
truncation. They also contain further information about the BPS states of the coset.

\section{Higher Spin Theories with $D(2,1| \alpha)$ Symmetry}\label{sec:hspin}

As a first step towards obtaining a (${\cal W}$-extended) large ${\cal N}=4$ asymptotic symmetry in a classical Vasiliev 
higher spin theory on AdS$_3$, we need to have a higher spin algebra based on the global 
$D(2,1| \alpha)$ subalgebra. We will now show that this can be 
achieved using the conventional oscillator construction of the supersymmetric higher spin algebra but now enhanced with 
a ${\rm U}(2)$ Chan-Paton index. This algebra can then be used to construct a Vasiliev set of equations for higher spin 
fields coupled to massive matter fields. A brief introduction to $D(2,1|\alpha)$ can be found in Appendix~\ref{app:D21}.

\subsection{Realising the ${\cal N}=2$ higher spin algebra}

Let us begin by briefly recalling how the ${\cal N}=2$ supersymmetric higher spin superalgebra  ${\rm shs}[\mu]$ \cite{BdWV}
is constructed --- this is the basis for the ${\cal N}=2$ Vasiliev set of equations. 
We consider the algebra
\be\label{sBmu}
sB[\mu] = \frac{U(\mathfrak{osp}(1|2))}{\langle 
C^{\mathfrak{osp}} - \frac{1}{4} \mu (\mu-1) {\bf 1}\rangle} \ ,
\ee
which can be described in terms of the oscillators $\hat{y}_\alpha$, $\alpha=1,2$, and $k$, subject to the relations
\cite{Prokushkin:1998bq,Prokushkin:1998vn}
\be\label{osccom}
[\hat{y}_\alpha,\hat{y}_\beta] = 2 i \epsilon_{\alpha\beta} (1 + \nu k) \ , \qquad
k \, \hat{y}_\alpha = - \hat{y}_\alpha k \ , \qquad k^2 = 1 \ .
\ee
Here $\nu=2\mu-1$, and $\epsilon_{12}=-\epsilon_{21}=1$ is antisymmetric. We can turn $sB[\mu]$ into a super Lie algebra
by defining (anti)-commutators as $[A,B]_{\pm}:= A\star B \pm B\star A$, where $\star$ is the 
associative product in $sB[\mu]$; as a super Lie algebra we then have 
\be
sB[\mu] = {\rm shs}[\mu] \oplus \mathbb{C} \ , 
\ee
where $\mathbb{C}$ corresponds to the unit generator ${\bf 1}$ of $U(\mathfrak{osp}(1|2))$. 
A basis for $sB[\mu]$ can be described by 
\begin{equation} \label{shs}
V_m^{(s)\pm} \sim \hat{y}_{(\alpha_1...}\, \hat{y}_{\alpha_n)}(\mathbf{1} \pm k)\ , 
\end{equation}
where $V_m^{(s)\pm}$ has `spin' $s=1+\frac{n}{2}$ with $n\geq 0$, and $m$ takes the values  
$2m=N_1-N_2$, with $N_{1,2}$ being the number of $\hat{y}_{1,2}$; thus $m$ lies in the range
$-s+1\leq m \leq s-1$. The super Lie algebra ${\rm shs}[\mu]$ is then also generated by these
modes, except that  the two $s=1$ modes are proportional to one another, $V_m^{(1)\pm}\equiv \pm k$.
(Here we have used that the ${\bf 1}$ generator is not part of shs$[\mu]$.) Thus we have 2 sets of generators for each spin $s=\tfrac{3}{2}, 2, \tfrac{5}{2}, \ldots$ and one for $s=1$.

The super Lie algebra ${\rm shs}[\mu]$ contains in particular the `wedge' algebra of the ${\cal N}=2$ 
superconformal algebra as its maximal finite dimensional subalgebra. This algebra is generated by the 
${\rm U}(1)$-current zero mode $J_0$ which we may take to equal (by adding to it a multiple of the
central ${\bf 1}$ operator)
\be
J_0 = - \tfrac{1}{2} (\nu + k) \ ,
\ee
the supercharges
\be
G^\pm_{1/2} = \tfrac{1}{2 \sqrt{2}}\, e^{-{\rm i}\pi/4}\, \hat{y}_1 (1\pm k) \ , \qquad
G^\pm_{-1/2} = \tfrac{1}{2 \sqrt{2}}\, e^{-{\rm i}\pi/4}\, \hat{y}_2 (1\pm k) \ ,
\ee
as well as the $\mathfrak{sl}(2)$ M\"obius generators 
\be\label{Lnosc}
L_1 = \tfrac{1}{4i} \hat{y}_1 \hat{y}_1 \ , \qquad
L_0 = \tfrac{1}{8i} \bigl( \hat{y}_1 \hat{y}_2 +  \hat{y}_2 \hat{y}_1 \bigr) \ , \qquad
L_{-1} = \tfrac{1}{4i} \hat{y}_2 \hat{y}_2 \ .
\ee
These generators satisfy indeed the ${\cal N}=2$ wedge algebra anti-commutation relations, 
\begin{eqnarray}
{}[L_m,L_n] & = & (m-n) L_{m+n} \ , \\
{} [L_m,G^\pm_r] & = & \bigl( \tfrac{m}{2}-r \bigr)  G^\pm_{m+r}  \label{spin32} \\
{} \{G^+_r,G^-_s\} & = & 2 L_{r+s} + (r-s) J_{r+s} \ , \\
{} \{G^+_r,G^+_s\} & = & \{G^-_r,G^-_s\} = 0 \ , 
\end{eqnarray}
and
\be
{}[J_0,L_n] = 0 \ , \qquad [J_0,G^\pm_r] = \pm G^\pm_{r} \  . 
\ee

Note that the other spin two generators, 
such as $\tilde{L}_1 = \tfrac{1}{4i} \hat{y}_1 \hat{y}_1k$, lead, upon taking commutators with the $G^{\pm}_r$,
to generators of higher spin. In turn
these then generate the full infinite-dimensional algebra; thus the ${\cal N}=2$ wedge algebra is generically
the largest finite-dimensional subalgebra of ${\rm shs}[\mu]$.

\subsection{Realising the $D(2,1|\alpha)$ Higher Spin Algebra}\label{sec:CP}

To realise the higher spin algebra which contains $D(2,1|\alpha)$ as a subalgebra, we 
could try and use an oscillator construction for this superalgebra. It turns out to be simpler, however, to 
generalise the construction in the ${\cal N}=2$ case by 
introducing `Chan-Paton' indices. Thus we consider instead of $sB[\mu]$ the algebra
\be\label{CP1}
sB_M[\mu] \equiv sB[\mu] \otimes {\rm M}_M(\mathbb{C})  \ ,
\ee
i.e.\ the tensor product of $sB[\mu]$ with the algebra of $M\times M$ matrices. Obviously, 
$sB_M[\mu]$ is also an associative algebra, and it has a basis consisting of the pairs 
$(V^{(s)\pm}_{m},t^a)$, where $t^a$ with $a=1,\ldots, M^2$ is a basis for the $M\times M$ matrices.
The associated super Lie algebra ${\rm shs}_M[\mu]$ that is defined via 
\be\label{factor}
sB_M[\mu] = {\rm shs}_M[\mu]  \oplus \mathbb{C} 
\ee
has then $2M^2$ generators for each spin $s=\tfrac{3}{2},2,\tfrac{5}{2},\ldots$, as well
as $2M^2-1$ generators of spin $s=1$. As before, we have removed the identity element from
the algebra since it is central and does not appear in (anti-)commutators; only the generator
$J_0\otimes {\bf 1}_M$ can be generated in anti-commutators. The remaining $2M^2-2$ 
generators of spin $s=1$ then realise the Lie algebra 
$\mathfrak{sl}(M)\oplus \mathfrak{sl}(M)$. 
\smallskip

While the above construction is general, for realising $D(2,1|\alpha)$ we will focus on the case $M=2$. We shall take the  `gravity' $\mathfrak{sl}(2)$  to be given by 
$L_n\otimes {\bf 1}_2$, where $n=0,\pm1$, $L_n$ is given in (\ref{Lnosc}) and ${\bf 1}_2$ is the $2\times 2$ identity matrix in ${\rm M}_2(\mathbb{C})$.
We can then classify the remaining generators according to their `spin'. At spin $s=1$, we have the
generators
\be
A^{\pm,i} = \frac{1}{2} (1\pm k) \otimes \sigma_i \ , 
\ee
where the $\sigma_i$ run over the Pauli matrices; they form the Lie algebra 
$\mathfrak{su}(2)\oplus \mathfrak{su}(2)$. In addition, shs$_2[\mu]$ contains the spin $s=1$ generator $J_0\otimes {\bf 1}_2$.
At spin $s=\tfrac{3}{2}$, the eight generators are 
\be\label{hsGgen}
G^{\pm,\alpha\beta}_{1/2} = \frac{1}{2}  \hat{y}_1 (1\pm k) \otimes E_{\alpha\beta}\ , \qquad
G^{\pm,\alpha\beta}_{-1/2} = \frac{1}{2} \hat{y}_2 (1\pm k) \otimes E_{\alpha\beta}\ ,
\ee
where $E_{\alpha\beta}$ is the matrix whose only non-zero entry (equal to $1$) is in the $\alpha,\beta$ position. 
With respect to the
`gravity' $\mathfrak{sl}(2)$, they still satisfy individually (i.e.\ for fixed $\pm,\alpha\beta$) (\ref{spin32}), which just means that 
these generators are really of spin $s=\tfrac{3}{2}$. With respect to the two commuting $\mathfrak{su}(2)$'s on the other hand,
they transform, for each fixed $\pm$, in the $({\bf 2},{\bf 2})$. For example, one has
\be
{}[A^{+,j},G^{+,\alpha\beta}_{r}] = - \frac{1}{2} \hat{y}_\alpha (1+k) \otimes (E_{\alpha\beta}\, \sigma_j) \ , 
\ee
where $\alpha\equiv \alpha(r) =\tfrac{3}{2}-r$. Thus the 
$A^{+}$ generators act by matrix multiplication from the right, while for the $A^{-}$ generators we have instead
\be
{}[A^{-,j},G^{+,\alpha\beta}_{r}] = \frac{1}{2} \hat{y}_\alpha (1+k) \otimes (\sigma_j E_{\alpha\beta})  \ , 
\ee
i.e.\ they act from the left. Thus the generators $G^{+,*}_{r}$ sit in a $({\bf 2},{\bf 2})$ representation with respect
to the two commuting $\mathfrak{su}(2)$ algebras. The situation for the $G^{-,*}_{r}$ supercharges is similar, although the
roles of $A^\pm$ are now interchanged, i.e.\ we have 
\be
{}[A^{+,j},G^{-,\alpha\beta}_{r}] =  \frac{1}{2} \hat{y}_\alpha (1-k) \otimes  \sigma_j E_{\alpha\beta} \ , \qquad
{}[A^{-,j},G^{-,\alpha\beta}_{r}] = - \frac{1}{2} \hat{y}_\alpha (1-k) \otimes  E_{\alpha\beta} \,\sigma_j \ .
\ee
A similar analysis can be done for all the higher spin generators as well, and one finds that the fermionic generators
(half-integer spin) all transform as $({\bf 2},{\bf 2}) \oplus ({\bf 2},{\bf 2})$, while the bosonic generators (integer spin 
with $s\geq 2$)
transform as $({\bf 3},{\bf 1}) \oplus ({\bf 1},{\bf 3}) \oplus 2\cdot ({\bf 1},{\bf 1})$.

\medskip

Next we want to show that the Lie superalgebra $D(2,1|\alpha)$  is the maximal finite dimensional subalgebra of ${\rm shs}_2[\mu]$. Here the parameter $\alpha$ is related 
to $\mu$ (or equivalently $\nu$) via 
\be
\alpha =\frac{1+\nu}{1-\nu} =  \frac{\mu}{1-\mu} \ .
\ee
Recall from (\ref{D21alpha}) that $D(2,1|\alpha)$ is generated by the `gravity' $\mathfrak{sl}(2)$, two 
commuting $\mathfrak{su}(2)$ algebras, as well as $4$ supercharges (that transform in the $({\bf 2},{\bf 2})$ 
with respect to the two $\mathfrak{su}(2)$ algebras). We have already identified the 
$\mathfrak{sl}(2)\oplus \mathfrak{su}(2)\oplus\mathfrak{su}(2)$ algebra in ${\rm shs}_2[\mu]$, but it remains to find the 
$4$ suitable linear combinations of the $8$ supercharges $G^{\pm,\alpha\beta}$ that form the generators of 
$D(2,1|\alpha)$. We define the four generators
\be
G^{++}_{r} =e^{\pi {\rm i}/4} \, \hat{y}_\alpha k \otimes \left( \begin{matrix} 0 & 1 \cr 0 & 0 \end{matrix} \right) \ , \qquad
G^{--}_{r} = -e^{\pi {\rm i}/4} \,  \hat{y}_\alpha k \otimes \left( \begin{matrix} 0 & 0 \cr 1 & 0 \end{matrix} \right) \ ,
\ee
\be
G^{-+}_{r} = -\frac{e^{\pi {\rm i}/4}}{2}  \Bigl[ \hat{y}_\alpha \otimes {\bf 1}_{2} 
+\hat{y}_\alpha k \otimes  \left( \begin{matrix} 1 & 0 \cr 0 & -1 \end{matrix} \right) \Bigr] \ ,
\ \
G^{+-}_{r} = \frac{e^{\pi {\rm i}/4}}{2} \Bigl[ \hat{y}_\alpha \otimes {\bf 1}_{2} 
- \hat{y}_\alpha k \otimes  \left( \begin{matrix} 1 & 0 \cr 0 & -1 \end{matrix} \right) \Bigr]\ ,
\ee
where in each case $\alpha\equiv \alpha(r) =\tfrac{3}{2}-r$. These particular linear combinations transform in the $({\bf 2},{\bf 2})$ of 
the two $\mathfrak{su}(2)$ algebras. Furthermore, their anti-commutators have the form
\begin{eqnarray}
{} \{ G^{++}_{r} , G^{++}_s\} & = & \{ G^{+-}_{r} , G^{+-}_s\} = \{ G^{-+}_{r} , G^{-+}_s\} = 
\{ G^{--}_{r} , G^{--}_s\} = 0  \\[4pt]
{} \{ G^{++}_{r} , G^{+-}_s\} & = & 
 2 (r-s) \, (1+\nu)\, A^{++}_{r+s} 
\\[4pt]
{} \{ G^{++}_{r} , G^{-+}_s\} & = &  2  (r-s) \, (1-\nu)\, A^{-+}_{r+s}
\\[4pt]
{} \{ G^{++}_{r} , G^{--}_s\} & = & 
- 4  L_{r+s} - 2 \, (r-s)\, \bigl[ (1+\nu) A^{+3}_{r+s} + (1-\nu) A^{-3}_{r+s} \bigr]  \\[4pt]
{} \{ G^{+-}_{r} , G^{-+}_s\} & = & 
 4 L_{r+s}  + 2 \, (r-s)\, \bigl[ (1+\nu) A^{+3}_{r+s} - (1-\nu) A^{-3}_{r+s} \bigr]
\end{eqnarray}
\begin{eqnarray}
{}\{G^{+-}_{r},G^{--}_{s} \} & = & 
- 2  (r-s) \, (1-\nu) A^{--}_{r+s}  \\[4pt]
{}\{G^{-+}_{r},G^{--}_{s} \} & = &  - 2  (r-s) \, (1+\nu) A^{+-}_{r+s} \ ,
\end{eqnarray}
where the complex basis for the current generators $A^{\pm \pm}$ was introduced in Appendix~\ref{app:comp}.
These anti-commutation relations agree precisely with those of  $D(2,1|\alpha)$, 
see eq.~(\ref{GGpm}),  provided we identify
\be
(1+\nu) = 2 \gamma \ , \qquad (1-\nu) = 2 (1-\gamma) \ , 
\ee
i.e.\ 
\be\label{parrel}
\nu = 2\gamma -1 \ , \qquad \gamma = \frac{1}{2} (1+\nu)= \mu  \qquad \hbox{with} \quad
\alpha = \frac{\gamma}{1-\gamma} = \frac{1+\nu}{1-\nu} = \frac{\mu}{1-\mu}\ .
\ee
Here we have used the relation $\nu=2\mu -1$. Thus we have shown that ${\rm shs}_2[\mu]$ contains 
indeed $D(2,1|\alpha)$ as a subalgebra. 

It is instructive to decompose ${\rm shs}_2[\mu]$ into representations of $D(2,1|\alpha)$ as 
\be\label{D2mult}
{\rm shs}_2[\mu] = D(2,1|\alpha) \oplus \bigoplus_{s=1}^{\infty} R^{(s)} \ , 
\ee
where $R^{(s)}$ is the $D(2,1|\alpha)$ multiplet consisting of the fields 
\be
\begin{array}{lll}\label{D2mult2}
& s: & ({\bf 1},{\bf 1})  \\
& s+\tfrac{1}{2}: & ({\bf 2},{\bf 2})  \\
R^{(s)}: \qquad & s+1: & ({\bf 3},{\bf 1}) \oplus ({\bf 1},{\bf 3})  \\
& s+\tfrac{3}{2}: & ({\bf 2},{\bf 2})  \\
& s+2: & ({\bf 1},{\bf 1}) \ .
\end{array}
\ee
In particular, we observe that the first non-trivial multiplet $R^{(1)}$ (whose lowest spin $s=1$ component is
precisely $J_0\otimes {\bf 1}$) contains a field of spin $s=3$, and 
thus will generate (upon taking (anti-)commutators with itself, as well as with $D(2,1|\alpha)$) the full algebra. 
Thus, at least for generic values of $\mu$, $D(2,1|\alpha)$ is the largest finite dimensional subalgebra of ${\rm shs}_2[\mu]$. 

We should mention that the $\mathfrak{u}(1)$ generator of the higher spin algebra $J_0\otimes {\bf 1}$ commutes
with all bosonic higher spin currents, while it has eigenvalues $\pm 1$ on the fermionic currents in the 
$({\bf 2},{\bf 2}) \oplus ({\bf 2},{\bf 2})$ --- in fact, since $J_0\otimes {\bf 1}$ is not part of the $D(2,1|\alpha)$ algebra, 
the commutator with $J_0\otimes {\bf 1}$ exchanges the spin $s+\tfrac{3}{2}$ generators of $R^{(s)}$ and $R^{(s+1)}$
with one another.

We also note in passing that the higher spin superalgebras
of the form ${\rm shs}({\cal N}|2)$, which are based on the $\mathfrak{osp}({\cal N}|2)$ global superalgebra, appear 
not to admit the oscillator deformation parameter $\nu$ (or equivalently $\mu$ is fixed to be $\tfrac{1}{2}$) when the 
number of supersymmetries ${\cal N} >2$ \cite{Henneaux:2012ny}. Thus the above construction of the higher spin 
superalgebra based on $D(2,1|\alpha)$ gives a way to  have a one parameter family of higher
spin algebras with extended supersymmetry ${\cal N}>2$.

\subsection{The Vasiliev Higher Spin Theory}

We can use the ${\rm shs}_2[\mu]$ algebra to construct a higher spin theory on AdS$_3$. The advantage of using the 
Chan-Paton construction is that the generalisation is straightforward. We know that the Vasiliev equations can be 
generalised to one in which the basic dynamical fields $W, S$ and $B$ do not just belong to the higher spin algebra 
${\rm shs}[\mu]$, but are also $M\times M$ matrices \cite{Prokushkin:1998bq, Prokushkin:1998vn}. Thus, in particular, 
one can consider the case of $M=2$ and hence view the fields as taking values in ${\rm shs}_2[\mu]$. To go from the 
complex Lie algebra to the real $\mathfrak{u}(2)$ algebra for the fields, we need to impose an appropriate reality 
condition on the fields. This consists of the usual self adjointness condition on the matrix sector of the fields, together 
with an involution of the higher spin algebra defined for the fields $W$, $S$, and $B$ in 
\cite{Prokushkin:1998bq, Prokushkin:1998vn}. 

The field $W$ contains the higher spin gauge fields and as mentioned earlier we have $2M^2-1=7$ spin one fields 
(two sets of $\mathfrak{su}(2)$ gauge fields, together with a $\mathfrak{u}(1)$) as well as $2M^2=8$ fields of spin 
$s=\tfrac{3}{2}, 2, \tfrac{5}{2} \ldots$. 
Note that there is a distinguished spin two field which corresponds to the $\mathfrak{sl}(2)$ in the global part of the 
higher spin algebra. The field $S$ is entirely auxiliary. The field $B$ in three dimensions is essentially auxiliary 
except for its lowest modes. 

\subsection{The Fundamental Representations of {shs$_2[\mu]$}}\label{sec:fund}

In order to describe the spectrum of the scalar fields, which are the lowest components of $B$, 
let us first review the situation for the higher
spin theory based on ${\rm shs}[\mu]$. In that case the scalar fields correspond to the fundamental representations 
of ${\rm shs}[\mu]$, i.e.\
to the representations of $\mathfrak{osp}(1|2)$ with $C^{\mathfrak{osp}}=\tfrac{1}{4}\mu (\mu-1)$. In terms of the
oscillator formulation of ${\rm shs}[\mu]$, the highest weight state $\phi$ of such a representation is annihilated 
by $\hat{y}_1$, and hence  has $L_0$ eigenvalue 
\be
L_0 \, \phi = \frac{1}{4} (1+\nu k) \phi \ , 
\ee
as follows directly from the definition of $L_0$ in (\ref{Lnosc}) together with (\ref{osccom}). Depending on 
the sign of the $k$ eigenvalue, $k\phi_\pm = \pm \phi_\pm$, we therefore have the $L_0$ eigenvalues 
\be
h_+ = \frac{1}{4} (1+\nu) = \frac{\mu}{2} \ , \qquad \qquad h_- = \frac{1}{4} (1-\nu) = \frac{1}{2} (1-\mu) \ .
\ee
The corresponding mass of the scalar field is then $M^2=\Delta (\Delta-2)$ where $\Delta=2h$, i.e. 
\be\label{scalmass}
M_+^2  = -1 + (1-\mu)^2 \ , \qquad \qquad M_-^2= - 1+ \mu^2 \ . 
\ee
These representations are `short' representations of the ${\rm shs}[\mu]$ algebra, i.e.\ they have a null-vector
\be
G^\mp_{-1/2}  \phi_\pm = 0 \ ,
\ee
but also a non-trivial fermionic descendant $G^{\pm}_{-1/2}\phi_\pm \neq 0$, which gives rise to a Dirac fermion
of mass $m^2=(\Delta-1)^2 = (\mu-\tfrac{1}{2})^2$. Their character therefore equals
\be\label{N2char}
{\rm Tr}_{R_{\pm}} (q^{L_0}) = q^{h_\pm} \, \frac{(1+q^{1/2})}{(1-q)} \ . 
\ee
\medskip

The fundamental representations of shs$_2[\mu]$ can be constructed similarly by taking the tensor product
of a fundamental representations of shs$[\mu]$ together with the defining $2$-dimensional representation
of the $M_2(\mathbb{C})$ matrix algebra. The highest weight state of the resulting representation is then not only 
annihilated by all positive modes, i.e.\ by $\hat{y}_1$, but also by $A^{\pm +}$, i.e.\ it is the `top' component
of the $2$-dimensional representation space. This state has then $L_0$ eigenvalues $h_\pm$, corresponding to 
a scalar field of mass $M_\pm$, but there is now a doublet of such states (corresponding to the $2$-dimensional
auxiliary space). The representations are `short', i.e.\ there are the null states
\be\label{Gshort}
G^{+\ast}_{-1/2} \phi_+ = 0 \ , \qquad G^{\ast+}_{-1/2} \phi_- = 0 \ , 
\ee
but the other descendants do not vanish, i.e.\ there is also a doublet of fermionic descendants. Their quantum numbers
with respect to the two $\mathfrak{su}(2)$ algebras are therefore
\be\label{matqno}
\phi_+ :  ({\bf 2}, {\bf 1})_0 \oplus ({\bf 1},{\bf 2})_{1/2}  \ , \qquad\qquad  \phi_-: ({\bf 1},{\bf 2})_0 \oplus  ({\bf 2}, {\bf 1})_{1/2}\ , 
\ee
where the first (second) quantum number refers to the $A^{+i}$ and $A^{-i}$ algebra, respectively, and the
index $(0,1/2)$ denotes the ground states or the first excited states, respectively. The specialised character is then
simply twice that of (\ref{N2char}). From the AdS point of view, these representations describe 
propagating modes corresponding to two complex massive scalars and two Dirac fermions; 
the mass of the Dirac fermions is always $m^2= (\mu-\tfrac{1}{2})^2$, while the scalars have mass $M^2=-1+(1-\mu)^2$
for the case of $R_+$, and mass $M^2=-1+\mu^2$ for the case of $R_-$. 
The full classical equations of motion for these matter fields are the matrix generalisations of the ones 
given in  \cite{Prokushkin:1998bq, Prokushkin:1998vn}. 

\subsection{Asymptotic Symmetry Algebra}\label{sec:asymp}

The asymptotic symmetries of the Vasiliev theory are much larger than those of  the higher spin algebra (i.e.\ 
shs$_2[\mu]$ in our case), as has been appreciated in the last few years \cite{Henneaux:2010xg, Campoleoni:2010zq, Gaberdiel:2011wb, Campoleoni:2011hg, Henneaux:2012ny, Hanaki:2012yf}. In fact, the asymptotic symmetry algebra can simply be obtained by performing the 
algebraic Drinfeld-Sokolov reduction of the higher spin algebra \cite{Gaberdiel:2011wb,Campoleoni:2011hg}.
The resulting classical algebra, which we shall denote by $s\tilde{\cal W}^{(4)\, {\rm cl}}_{\infty}[\mu]$, is generated by the same sort
of modes as shs$_2[\mu]$, except that the `wedge' condition is relaxed. 
In other words, it will have a 
basis labelled by $(V^{(s)\pm}_{m},t^a)$, where the $t^a$ form a basis for ${\rm U}(2)$ as before, but $m$ is now no longer 
restricted by the condition $|m| <s$ (but rather lies in $m\in \mathbb{Z} + s$).
The structure constants of the non-linear algebra obeyed by these generators 
are largely determined by the requirement that $s\tilde{\cal W}^{(4)\, {\rm cl}}_{\infty}[\mu]$
is a Poisson algebra, satisfying the usual Jacobi identities.

Since shs$_2[\mu]$ contains $D(2,1|\alpha)$ with $\gamma=\mu$ as a proper subalgebra, the classical super algebra 
$s\tilde{\cal W}^{(4)\, {\rm cl}}_{\infty}[\mu]$ will contain the (classical version of the) 
non-linear large ${\cal N}=4$ superconformal algebra $\tilde{A}_\gamma$ as a subalgebra --- this just follows from the fact 
that the wedge algebra of $\tilde{A}_\gamma$ 
is $D(2,1|\alpha)$.\footnote{Incidentally, the wedge algebra of the large ${\cal N}=4$ superconformal
algebra $A_\gamma$ itself is also $D(2,1|\alpha)$ (together with a central generator), so one may have thought that 
$s{\cal W}^{(4)\, {\rm cl}}_{\infty}[\mu]$ 
should contain $A_\gamma$, rather than its non-linear truncation $\tilde{A}_\gamma$. However, $A_\gamma$ contains
in particular four free fermion generators, that cannot appear from the asymptotic symmetry analysis
based on shs$_2[\mu]$, and thus this possibility is excluded. The large ${\cal N}=4$ algebra $A_\gamma$
as well as its non-linear truncation $\tilde{A}_\gamma$ are discussed in detail in Appendix~\ref{app:alg}.} However, unlike the situations 
that were previously studied, in our case the structure of the resulting $s\tilde{\cal W}^{(4)\, {\rm cl}}_{\infty}[\mu]$ algebra differs 
quite significantly from that of shs$_2[\mu]$. One instance that illustrates this phenomenon is the following.
Consider the higher spin algebra shs$_2[\mu]$ for the case when
$\mu=N+1$ with $N$ a positive integer. Then, the underlying shs$[N+1]$ algebra can be truncated to 
$\mathfrak{sl}(N+1|N)$, and the decomposition (\ref{D2mult}) terminates as 
\be\label{shstrunc}
{\rm shs}_2[N+1] = D(2,1| - \tfrac{N+1}{N}) \oplus \bigoplus_{s=1}^{N-1} R^{(s)} \ \oplus \hat{R}^{(N)}_- \ ,
\ee
where $\hat{R}^{(N)}_-$ is the short representation of $D(2,1| - \tfrac{N+1}{N})$ with spin content
\be
\begin{array}{lll}\label{D2multshort}
& N: & \quad ({\bf 1},{\bf 1})  \\
\hat{R}^{(N)}_-: \qquad  & N+\tfrac{1}{2}: & \quad ({\bf 2},{\bf 2})  \\
& N+1: & \quad ({\bf 1},{\bf 3})  \ . 
\end{array}
\ee
(There is a similar representation $\hat{R}^{(N)}_+$ where instead of the   $({\bf 1},{\bf 3})$ representation
the  $({\bf 3},{\bf 1})$ is retained at spin $N+1$; this representation appears at $\alpha=-\tfrac{N}{N+1}$, 
reflecting that under the exchange $\alpha \mapsto \alpha^{-1}$ the roles of the two $\mathfrak{su}(2)$
algebras are interchanged. These representations are discussed in more detail in Appendix~\ref{app:D21}.) 

One would therefore expect that the corresponding $s\tilde{\cal W}^{(4)\, {\rm cl}}_{\infty}[\mu]$ algebra is similarly truncated, i.e.\
that it is generated by the spin content described by (\ref{shstrunc}). However, this is not the case. The reason
is that, unlike $D(2,1|\alpha)$, the large ${\cal N}=4$ algebra $\tilde{A}_\gamma$  (or indeed its classical version) 
does {\em not}
possess a short representation of the form $\hat{R}^{(N)}_{-}$. Recall from Appendix~\ref{app:D21}
that the ideal by which one
has to quotient $R^{(N)}$ in order to obtain $\hat{R}^{(N)}_{-}$ is generated by ${\cal N}$ in (\ref{idealgen})
with $s=N$. However, for $\tilde{A}_\gamma$ this vector does not generate an ideal since 
\be
A^{+-}_{1} \, G^{+-}_{-\frac{1}{2}} \, G^{++}_{-\frac{1}{2}} \, \Phi_N = 
G^{--}_{\frac{1}{2}}\, G^{++}_{-\frac{1}{2}} \, \Phi_N =  - 
4 L_0 \Phi_N = - 4 N \Phi_N \neq 0 \ .
\ee
(Note that the relevant generator $A^{+-}_{1}$ is not part of $D(2,1|\alpha)$, and hence this constraint
is invisible from the point of view $D(2,1|\alpha)$.) Thus it is impossible to truncate $s\tilde{\cal W}^{(4)\, {\rm cl}}_{\infty}[\mu]$
in this manner.

So far we have discussed the classical $s\tilde{\cal W}^{(4)\, {\rm cl}}_{\infty}[\mu]$ algebra; at finite central charge one
expects further corrections to the structure constants that arise from normal ordering terms, see 
\cite{Gaberdiel:2012ku} for a detailed explanation of this phenomenon. The resulting `quantum' algebra
$s\tilde{\cal W}^{(4)}_{\infty}[\mu]$ should then be equivalent to the coset algebras that will be discussed in the following
section. If this is indeed true, then $s\tilde{\cal W}^{(4)}_{\infty}[\mu]$ must truncate to finitely generated ${\cal W}$-algebras
at least for certain rational values of $\mu$ at the appropriate value of the central charge; however, as for the case 
discussed in \cite{Gaberdiel:2012ku}, this truncation phenomenon is unlikely to be visible from the point of view of the 
classical $s\tilde{\cal W}^{(4)\, {\rm cl}}_{\infty}[\mu]$ algebra.

One would also expect that the quantum algebra $s\tilde{\cal W}^{(4)}_{\infty}[\mu]$ should exhibit some sort of triality 
identifications as in \cite{Gaberdiel:2012ku} (or as in \cite{Candu:2012tr} for ${\cal N}=2$). However, as explained in Appendix~\ref{app:nonl}, since the levels of the affine $\mathfrak{su}(2)$
algebras appear explicitly in $s\tilde{\cal W}^{(4)}_{\infty}[\mu]$, the only non-trivial relation is the symmetry
$\gamma\leftrightarrow 1-\gamma$ that is already visible at the classical level. The fact that we have no non-trivial
relation between integer $\mu$ and fractional $\mu$ 
is compatible with the fact that $s\tilde{\cal W}^{(4)}_{\infty}[\mu]$ does not truncate at integer $\mu$ --- in fact, 
the reason why such a relation had to exist for the cases discussed in \cite{Gaberdiel:2012ku,Candu:2012tr}
was that both algebras in question had the same spin content, and hence had to agree for some suitable
identification of $\mu$.

\section{Large ${\cal N}=4$ Cosets}\label{sec:coset}

Next we want to identify candidate 2d conformal field theories which might be dual, in the large $N$ limit, 
to the bulk Vasiliev higher spin theories containing the global $D(2,1|\alpha)$ superalgebra.
As in the cases with smaller supersymmetry \cite{Creutzig:2011fe, Candu:2012jq} we might expect the dual to be a 
coset CFT as well.  Coset theories with large ${\cal N}=4$ superconformal symmetry 
have not been systematically explored or classified unlike, say, 
the ${\cal N}=2$ theories that were analysed in detail by 
Kazama and Suzuki \cite{Kazama:1988qp, Kazama:1988uz}. 

However, there are some coset theories that are expected to possess the large ${\cal N}=4$ superconformal symmetry
\cite{Sevrin:1988ew,Schoutens:1988ig}.
These are in particular the cosets based on Wolf symmetric spaces such as 
$\frac{{\rm SU}(N+2)}{{\rm SU}(N)\times {\rm U}(1)}$  
\cite{Spindel:1988sr,Goddard:1988wv,VanProeyen:1989me,Sevrin:1989ce}, see
also \cite{Gates:1995ip} for subsequent developments.
This motivates one to look in more detail at the cosets
\be\label{N41}
\frac{\mathfrak{su}(N+2)^{(1)}_{\kappa}}{\mathfrak{su}(N)^{(1)}_{\kappa}}  \cong \frac{\mathfrak{su}(N+2)_{k} 
\oplus \mathfrak{so}(4N+4)_1}{\mathfrak{su}(N)_{k+2} }  \ ,
\ee
where the superscript `$(1)$' denotes the ${\cal N}=1$ superconformal affine algebra, and the level $\kappa$
on the left-hand side equals 
$\kappa=k+N+2$. Here the denominator is embedded into the numerator in the standard fashion, i.e.\ 
in terms of matrices, the $\mathfrak{su}(N)$ of the denominator is the first $N\times N$ block of the 
$(N+2)\times (N+2)$ matrix of the numerator. In going to the bosonic description on the right-hand-side we have
used that (see Section~\ref{sec:currents} for a brief review)
\be\label{susybos}
\mathfrak{g}_\kappa^{(1)} \ \cong \ \mathfrak{g}_{\kappa - h^\vee} \oplus (\dim(\mathfrak{g})\ \hbox{free fermions}) \ ,
\ee
where $h^\vee$ is the dual Coxeter number of $\mathfrak{g}$, and we have employed the fact that 
$d$ free fermions generate $\mathfrak{so}(d)_1$. 

As we will see in more detail below, we will actually be considering a slightly different coset, namely,  
\be\label{N4coset2}
\frac{\mathfrak{su}(N+2)^{(1)}_{\kappa} }{\mathfrak{su}(N)^{(1)}_{\kappa} \oplus \mathfrak{u}(1)}
\oplus \mathfrak{u}(1) \ .
\ee
This will make a difference for the identification of the ${\rm U}(1)$ charges and conformal dimensions, 
but not materially affect the construction of the other generators of the algebra. 
In a final
step we will also divide out $4$ free fermions as well as the $\mathfrak{u}(1)$ factor in the numerator
to go to the non-linear form (that contains $\tilde{A}_\gamma$ rather than $A_\gamma$ as a subalgebra). 
However, also this final step has a rather minimal effect on most aspects of our discussion, and thus 
for many purposes we will continue to work with the simpler form in (\ref{N41}).

The central charge of the coset (\ref{N41}) or indeed (\ref{N4coset2}), computed as the difference between 
the numerator and denominator WZW theories, equals
\begin{equation}\label{cN4k}
c_{N,k} = \frac{6 (k+1) (N+1)}{k+N+2} \ .
\end{equation}
This agrees with the general form of the central charge of the large ${\cal N}=4$ algebra $A_\gamma$, see 
eq.~(\ref{A9}), for 
\begin{equation}\label{kpkm}
k^+ = (k+1) \ , \qquad k^- = (N+1) \ . 
\end{equation}
It was shown in \cite{VanProeyen:1989me,Sevrin:1989ce} that the coset (\ref{N41}) contains indeed $A_\gamma$, 
and thus (\ref{cN4k}) is very suggestive. The parameter $\gamma = \frac{k^-}{k^++k^-}$ of the large ${\cal N}=4$ algebra
then takes the value 
\be\label{gamcos}
\gamma =\frac{N+1}{N+k+2} \ \ \Longrightarrow \ \ \alpha= \frac{\gamma}{1-\gamma}= \frac{N+1}{k+1}\ ,
\ee
where we have used (\ref{alphgam}).
In the following we shall identify the two commuting $\mathfrak{su}(2)$ algebras with levels (\ref{kpkm}). We shall 
also describe the other generators of the coset ${\cal W}$-algebra.

\subsection{Constructing the Two $\mathfrak{su}(2)$ Affine Algebras}\label{sec:currents}

We shall mainly work with the ${\cal N}=1$ superconformal affine algebra description on the left-hand-side
of (\ref{N41}), and thus we need to review the structure of these algebras.
The ${\cal N}=1$ superconformal algebra $\mathfrak{g}_\kappa^{(1)}$ is generated by the currents
${\cal J}^a$, satisfying a $\mathfrak{g}_\kappa$ affine algebra 
\be
{}[ {\cal J}^a_m, {\cal J}^b_n ] = i f^{abc} {\cal J}^c_{m+n} + \kappa \, \delta^{ab} \delta_{m,-n}  \ ,
\ee
as well as $\dim(\mathfrak{g})$ free fermions $\psi^a_r$, transforming in the adjoint representation 
of $\mathfrak{g}$, 
\begin{eqnarray}
{}[{\cal J}^a_m, \psi^b_r ] & = & i f^{abc}\, \psi^c_{m+r} \\
{}\{\psi^a_r,\psi^b_s\} & = & \delta^{ab}\, \delta_{r,-s} \ .
\end{eqnarray}
Given the $\dim(\mathfrak{g})$ free fermions $\psi^a_r$, we can construct an affine algebra
at level $k=h^\vee$ by 
\be\label{Mcurr}
M^a_n = \frac{i}{2} f^{abc}\, \sum_r  \psi^b_{n-r} \psi^c_{r} \ ,
\ee
with respect to which the free fermions transform also in the adjoint representation. Thus the currents
\be
{\cal J}^{(b)\, a}_m \equiv {\cal J}^a_m - M^a_m \ ,
\ee
commute with the free fermions, and hence with the current generators $M^a_n$. It follows that
the algebra generated by the ${\cal J}^{(b)\, a}_m$ is again an $\mathfrak{g}$ affine algebra, but now at level $k = \kappa-h^\vee$,
thus demonstrating (\ref{susybos}).
\medskip

Next we want to determine the spectrum of the ${\cal W}$-algebra generators. We begin by decomposing 
$\mathfrak{su}(N+2)$ into $\mathfrak{su}(N)$ representations as
\be\label{decompo}
\mathfrak{su}(N+2) = \mathfrak{su}(N) \oplus \mathfrak{su}(2) \oplus \mathfrak{u}(1) \oplus 
({\bf N},{\bf 2}) \oplus (\bar{\bf N},{\bf 2}) \ .
\ee
Thus the coset contains an $\mathfrak{su}(2)^{(1)}_\kappa$ affine algebra at level $\kappa=N+k+2$, whose generators
we shall denote by ${\cal J}^{a}$,  as well as 
a  $\mathfrak{u}(1)^{(1)}$ algebra. The other generators carry charge with respect to the $\mathfrak{su}(N)$ 
algebra of the denominator. In fact, since we are working with the ${\cal N}=1$ formulation,  we may take the other generators
to consist of fermions and bosons transforming as 
\be\label{cocur}
{\cal J}^{(b)\, i,\alpha}\ , \ \ \psi^{i,\alpha} \ : \ ({\bf N},{\bf 2}) \quad \qquad \hbox{and} \qquad \quad
\bar{\cal J}^{(b)\, i,\alpha}\ , \ \ \bar\psi^{i,\alpha} \ : \ (\bar{\bf N},{\bf 2}) \ , 
\ee
where $i\in\{1,\ldots,N\}$ denotes the vector index of the fundamental (or antifundamental) representation 
of $\mathfrak{su}(N)$,
while $\alpha\in\{1,2\}$ is the index of the 2-dimensional representation of $\mathfrak{su}(2)$. 

In the vacuum  representation (i.e.\ for the purpose of determining the ${\cal W}$-algebra), we are only interested in 
$\mathfrak{su}(N)$ singlet states. We can analyse these states for low conformal dimensions explicitly. Let us begin
by looking at the states at $h=1$. In addition to the currents coming from 
$\mathfrak{su}(2)^{(1)}_\kappa \oplus \mathfrak{u}(1)^{(1)}$, the only additional 
generators at $h=1$ can appear from bilinear singlets of the fermions, i.e.\ from the states
\be\label{Kcurrents}
\tilde{K}^{\alpha\beta} = \sum_{i=1}^{N} : \psi^{i,\alpha} \, \bar\psi^{i,\beta} : \ . 
\ee
They generate the affine algebra
\be
\mathfrak{su}(2)_N \oplus \mathfrak{u}(1) \ ,
\ee
where the level of $\mathfrak{su}(2)$ equals $N$. (This is obviously correct for $N=1$, and 
the general case is just the diagonal embedding into $N$ copies of the $N=1$ construction.)

It is easy to check that with respect to the currents (\ref{Kcurrents}), the free fermions $\psi^{i,\alpha}$ and
$\bar\psi^{j,\beta}$ transform, for each fixed $(i,j)$, in the ${\bf 2}$ of $\mathfrak{su}(2)$. Thus the generators 
\be
\tilde{J} = {\cal J} - \tilde{K} \ , 
\ee
where ${\cal J}$ denote the currents from $\mathfrak{su}(2)^{(1)}_\kappa$, 
commute with these free fermions, and hence with the currents (\ref{Kcurrents}). 
Thus we conclude that the ${\cal W}$-algebra contains the current algebras 
\be
[\tilde{J} \oplus \tilde{K} ]: \qquad \mathfrak{su}(2)_{k+2} \oplus \mathfrak{su}(2)_N \ .
\ee
This is still not quite what we want. The reason for this is that the $4$ free fermions that
are the fermionic generators of the $\mathfrak{su}(2)^{(1)}_\kappa \oplus \mathfrak{u}(1)^{(1)}$
algebra from above are singlets with respect to the algebra generated by the currents $\tilde{K}$ in (\ref{Kcurrents}),
whereas the free fermions $Q^a$ of the $A_\gamma$ algebra transform non-trivially with respect to both
$A^{\pm, i}$, see eq.\ (\ref{A2}). 

In order to correct this, we now first subtract out from the $\tilde{J}$-currents the $\mathfrak{su}(2)_2$ algebra
that is obtained by the $3$ free fermions in $\mathfrak{su}(2)^{(1)}_\kappa$ as in (\ref{Mcurr}); the resulting
currents $\hat{J}$ are then at level $k$, and commute with all $4$ free fermions. Out of these free fermions we 
then construct the current algebra
\be
\mathfrak{so}(4)_1 \cong \mathfrak{su}(2)_1 \oplus \mathfrak{su}(2)_1 \ ,
\ee
with respect to which the $4$ fermions transform as $({\bf 2},{\bf 2})$. We then add one $\mathfrak{su}(2)_1$
algebra to $\hat{J}$, and the other to $\tilde{K}$, and we denote the resulting generators by $J$ and $K$, respectively. 
The free fermions then transform in the $({\bf 2},{\bf 2})$ with respect to them. Furthermore, their levels are 
$k+1$ and $N+1$, as expected from (\ref{kpkm}).

\subsection{The Supercharges}

Next we consider the states at $h=\tfrac{3}{2}$. It is easy to see that we can construct eight 
$\mathfrak{su}(N)$ singlets at $h=\tfrac{3}{2}$, namely 
\be\label{8susy}
G^{\alpha\beta} = \sum_{i=1}^{N} : {\cal J}^{(b)\, i,\alpha} \, \bar\psi^{i,\beta}: \ , \qquad
\bar{G}^{\alpha\beta} = \sum_{i=1}^{N} : \bar{{\cal J}}^{(b)\, i,\alpha} \, \psi^{i,\beta}: \ , 
\ee
where we have used the same notation as in eq.~(\ref{cocur}). Both $G$ and $\bar{G}$ transform in the 
$({\bf 2},{\bf 2})$ with respect to the two  affine $\mathfrak{su}(2)$ algebras; these generators therefore 
mirror precisely the spin content of the higher spin algebra in eq.~(\ref{hsGgen}). 

We should note though that these generators do not directly define `supercharges'. Indeed, the 
actual supercharges of the large ${\cal N}=4$ algebra must have the property that their anticommutator
contains the full stress energy tensor of the theory. Since the supercharges in (\ref{8susy}) are nil-potent in 
the sense that 
\be
G^{\alpha\beta} G^{\gamma\delta} \sim {\cal O}(1) \ , \qquad
\bar{G}^{\alpha\beta} \bar{G}^{\gamma\delta} \sim {\cal O}(1)  \ , 
\ee
we need to combine the generators in order to form the actual supercharges. (Incidentally, this also
mirrors precisely what happens in the higher spin algebra analysis of Section~\ref{sec:CP}.) Furthermore, we 
need to correct them by composite fields of the form
\be
U \chi \ , \qquad  J \chi \ , \qquad K \chi \ , \qquad  \chi \chi \chi \ , 
\ee
where the $\chi\equiv \chi^{\alpha\beta}$ are the $4$ free fermions that transform in the $({\bf 2},{\bf 2})$
with respect to the two affine algebras , 
and $U$ is the $\mathfrak{u}(1)$ generator. In each case, one has to pick out the term that transforms in 
the $({\bf 2},{\bf 2})$.

\subsection{The Higher Spin Currents}\label{sec:hscur}

Next we want to describe the full spectrum of the ${\cal W}$-algebra. This can be done as in \cite{Candu:2012jq}.
Indeed, the character of the vacuum representation consists, for sufficiently large $k$ and $N$, of the 
$\mathfrak{su}(N)$ singlet states that can be formed out of the fermions and bosons in eq.~(\ref{cocur}). This
spectrum is generated by the fields that are bilinear in the generators of eq.~(\ref{cocur}) as well as 
their derivatives (but ignoring total derivaties). For example, the $\mathfrak{su}(N)$ singlets that can be formed
out of $\psi^{i,\alpha}$ and $\bar\psi^{i,\beta}$ and their derivatives, gives rise to four generating fields of each spin
$s=1,2,3,\ldots$. (The fields of spin $s=1$ are the currents $\tilde{K}$ we considered before.) These
fields transform in the ${\bf 1}\oplus {\bf 3}$ of the $\mathfrak{su}(2)$ algebra generated by the $K$-currents,
but are singlets with respect to the $\mathfrak{su}(2)$ algebra generated by the $J$-currents, as is clear 
from the structure of the two $\mathfrak{su}(2)$ algebras, see Section~\ref{sec:currents}.

Similarly, we get from the bilinears of the ${\cal J}^{(b)\, i,\alpha}$ and $\bar{\cal J}^{(b)\,i,\beta}$ four generating 
fields of each spin $s=2,3,\ldots$. They now transform in the ${\bf 1}\oplus {\bf 3}$ of the 
$\mathfrak{su}(2)$ algebra generated by the $J$-currents, but are singlets with respect to the $\mathfrak{su}(2)$ 
algebra generated by the $K$-currents. Finally, from the bilinears involving one 
fermion and one boson  we get $8$ generating fields of spin $s=\tfrac{3}{2},\tfrac{5}{2}, \tfrac{7}{2}, \ldots$. 
They transform in (two copies of) the $({\bf 2},{\bf 2})$ with respect to the two $\mathfrak{su}(2)$ algebras. 

Altogether the higher spin content of the coset theory therefore consists of $8$ higher spin fields of each spin 
$s=\tfrac{3}{2},2,\tfrac{5}{2},3,\ldots$, where the fermionic fields are in the $({\bf 2},{\bf 2})\oplus  ({\bf 2},{\bf 2})$, while 
the bosonic fields are in the $({\bf 1}\oplus{\bf 3},{\bf 1})\oplus  ({\bf 1},{\bf 1}\oplus {\bf 3})$. The resulting
${\cal W}$ algebra will be denoted by $s{\cal W}^{(4)}_{\infty}[\gamma]$. Its higher spin generators match
precisely those of the asymptotic symmetry algebra  of the higher spin theory based on 
${\rm shs}_2[\mu]$, see Sections~\ref{sec:CP} and \ref{sec:asymp}.

\subsection{The $\mathfrak{u}(1)$ Current}\label{sec:uone}

Finally, it is important to identify correctly the $\mathfrak{u}(1)$ generator of the resulting coset. 
The original coset (\ref{N41}) contains a natural $\mathfrak{u}(1)$ algebra, namely the one that 
appears in (\ref{decompo}). The corresponding 
generator is embedded as $\hat{U} = {\rm diag}(1,\ldots, 1, - \frac{N}{2}, -\frac{N}{2})$ into 
$\mathfrak{su}(N+2)$, and it precisely extends the $\mathfrak{su}(N)$ algebra of the denominator to
$\mathfrak{u}(N)$.  The `level' of this $\hat{U}$ generator is 
\begin{equation}
\hat\kappa = (k+N+2) \Bigl(N + 2 \cdot \frac{N^2}{4}\Bigr) = \frac{(k+N+2)}{2}\, N (N+2) \ . 
\end{equation}
With respect to $\hat{U}$, the free fermions and bosons 
$\psi^{i,\alpha}$, ${\cal J}^{(b)\, i,\alpha}$ and $\bar\psi^{j,\beta}$, $\bar{\cal J}^{(b)\, j,\beta}$
carry charge $\pm \tfrac{(N+2)}{2}$, respectively, while 
the $4$ free fermions $\chi^{\alpha\beta}$ are neutral. For reasons that will become clearer 
below when we study the representations of (\ref{N41}), this is however not the `correct' $\mathfrak{u}(1)$ algebra.
(Indeed, from a stringy point of view, the $\mathfrak{u}(1)$ generator should be related to the ${\rm S}^1$ of
the target space, and should therefore not be coupled directly to the ${\rm AdS}_3\times {\rm S}^3 \times {\rm S}^3$
part of the background.) Instead, as is implicit in (\ref{N4coset2}), it is much more natural to divide out by this $\hat{U}$ 
generator, and add in an additional independent $\mathfrak{u}(1)$ generator (that we shall denote by $U$). 
Incidentally, this is also in agreement with the Wolf symmetric space form for the cosets given 
in \cite{Sevrin:1989ce}.

\subsection{The Non-Linear ${\cal N}=4$ Algebra}\label{sec:nlin}

The coset theory we have described so far actually does not directly match with the higher spin theory
based on the algebra ${\rm shs}_2[\mu]$. Indeed, the coset algebra $s{\cal W}^{(4)}_{\infty}[\gamma]$ 
includes the large ${\cal N}=4$ algebra $A_\gamma$, and therefore contains $4$ free fermions as well as 
a $\mathfrak{u}(1)$ current 
--- the corresponding generators are denoted by $Q^a_r$ and $U_m$ in Appendix~\ref{app:alg}, respectively. On
the other hand, these generators are not visible in the $D(2,1|\alpha)$ wedge subalgebra, and therefore also
do not appear in the higher spin theory of Section~\ref{sec:hspin}. (Note that the higher spin theory contains
a $\mathfrak{u}(1)$ current, namely the bottom component of $R^{(1)}$, see eq.~(\ref{D2mult}). However, this is not
to be confused with the $\mathfrak{u}(1)$ generator of the $A_\gamma$ algebra; indeed, their 
transformation properties with respect to $D(2,1|\alpha)$ are different.)

There is however a standard way to remedy this problem. As was explained quite generally in 
\cite{Goddard:1988wv}, one can always factor out the free fermions and the $\mathfrak{u}(1)$ current
from the $A_\gamma$ algebra, and similarly therefore also from $s{\cal W}^{(4)}_{\infty}[\gamma]$.
The resulting algebra will be called $s\tilde{\cal W}^{(4)}_{\infty}[\gamma]$, and it then contains 
the non-linear $\tilde{A}_\gamma$ algebra as a subalgebra, see Section~\ref{app:nonl}. As is explained there,
the anti-commutator of the supercharges of $\tilde{A}_\gamma$ 
contains a term that is bilinear in the $\mathfrak{su}(2)\oplus \mathfrak{su}(2)$ currents, and 
the structure constants acquire $1/(k^++k^-)$ corrections. Apart
from that, however, rather little changes: in particular, the higher spin content is unaffected by this
procedure, while the central charge is just reduced by $3$, i.e.\ we have 
\be
c_{\rm non-linear} = \frac{ 6 k^+ k^-}{k^+ + k^-} - 3  = \frac{ 6 \hat{k}^+ \hat{k}^- +3(\hat{k}^+ + \hat{k}^-)}{\hat{k}^+ + \hat{k}^-+2 }  \ ,
\ee
where $\hat{k}^\pm = k^\pm -1$ are the levels of the $\mathfrak{su}(2)\oplus \mathfrak{su}(2)$ currents in $\tilde{A}_\gamma$.
With this modification, the spin spectrum of the higher spin theory and the coset theory then match precisely
in the 't~Hooft limit. In particular,
the $\mathfrak{u}(1)$ generator of the higher spin theory $(J_0 \otimes {\bf 1})$ can be identified with the $\mathfrak{u}(1)$
current coming from (\ref{Kcurrents}). Indeed, the zero mode of the latter commutes with all bosonic higher spin
currents of the coset, while it has eigenvalues $\pm 1$ on $G^{\alpha\beta}$ and $\bar{G}^{\alpha\beta}$,
and similarly for the fermionic higher spin currents. This therefore matches precisely what was found for
$(J_0\otimes {\bf 1})$ at the end of  Section~\ref{sec:CP}. 

In the following we shall mostly work with the $s{\cal W}^{(4)}_{\infty}[\gamma]$ algebra, containing the  large $A_\gamma$ 
algebra (rather than with $s\tilde{\cal W}^{(4)}_{\infty}[\gamma]$), 
since this is often more convenient. However, it is straightforward to convert results from $s{\cal W}^{(4)}_{\infty}[\gamma]$
to $s\tilde{\cal W}^{(4)}_{\infty}[\gamma]$ since the free fermions will not play any role in the following, and
the additional $\mathfrak{u}(1)$ generator will just go along for the ride.

\section{The Coset Spectrum}\label{sec:spec}

In this section we compute the dimensions of some of the primary representations of the coset 
(\ref{N4coset2}). We will be primarily interested in the BPS representations (which saturate the 
BPS bound of the large ${\cal N}=4$ $A_\gamma$ algebra, see appendix~\ref{app:bound}). 
Based on our sample calculations we will present the result for the full BPS spectrum in Section~\ref{sec:BPSgen}.
\smallskip

The conformal dimension of a coset primary can be easily calculated, using the conformal dimensions of
the mother and daughter theories. For example, for the case of the coset (\ref{N4coset2}), the representations
are labelled by an integrable highest weight representation $\Lambda_+$ of $\mathfrak{su}(N+2)_{k}$, an
integrable highest weight representation $\Lambda_-$ of $\mathfrak{su}(N)_{k+2}$, as well 
as the quantum numbers $u$ and $\hat{u}$ of the numerator and denominator $\mathfrak{u}(1)$ algebras.
The corresponding conformal dimension then equals
\be\label{hcos}
h(\Lambda_+;\Lambda_-)  = \frac{C^{(N+2)}(\Lambda_+)}{k+N+2} - 
\frac{C^{(N)}(\Lambda_-)}{k+N+2} - \frac{\hat{u}^2}{N(N+2)(N+k+2)} + \frac{u^2}{N+k+2} + n \ , 
\ee
where $C^{(L)}$ is the quadratic Casimir of $\mathfrak{su}(L)$,
and $n$ is the excitation number. (This excitation 
number may be integer or half-integer, since the ${\cal N}=1$ superconformal affine algebra also contains
free fermions.) Let us now illustrate this formula with a number of examples. Since the $\mathfrak{u}(1)$ generator in the numerator of the coset will have to be quotiented out in comparing to the higher spin theory of  Section~\ref{sec:hspin} (see the discussion in  Section~\ref{sec:nlin}), we shall always set $u=0$ in the following. We note that $u$
just goes along for the ride, i.e.\ it can be chosen independently, and it does not affect
the BPS condition, compare  eqs.~(\ref{hcos}) and (\ref{A28}). It is therefore consistent
to set it to zero, as expected on general grounds.

\subsection{The Minimal Representations}\label{sec:minrep}

The simplest representation to consider is the $(0;{\rm f})$ representation, i.e.\ the representation where
$\Lambda_+=0$ and $\Lambda_-={\rm f}$, the fundamental representation of 
$\mathfrak{su}(N)$. The corresponding states are of the form
\begin{equation}
\psi^{i,\alpha}_{-1/2} |0\rangle \ .
\end{equation}
They have $l^+=0$ and $l^-=\tfrac{1}{2}$, since the free fermions $\psi^{\alpha,i}$ transform in the ${\bf 2}$
with respect to $K$, but are singlets with respect to $J$. Furthermore, they carry $\mathfrak{u}(1)$ charge 
$\hat{u}=\frac{N+2}{2}$. Thus their conformal dimension equals
\begin{equation}\label{0f}
h(0;{\rm f}) = \frac{1}{2} - \frac{C^{(N)}({\rm f})}{k+N+2} - \frac{(N+2)^2}{4N (N+2) (N+k+2)}= \frac{k+\frac{3}{2}}{2 (k+N+2)} \ ,
\end{equation}
since $C^{(N)}({\rm f}) = \tfrac{N}{2} - \tfrac{1}{2N}$. This is now to be compared with the
BPS bound, eq.~(\ref{A28}), which takes the form 
\begin{eqnarray}
h(l^+=0,l^- = \tfrac{1}{2},u=0)_{\rm BPS} & = &  \frac{1}{(N+k+2)} \Bigl( \tfrac{1}{2}(k+1) + \tfrac{1}{4} \Bigr)  \\ 
& = & \frac{1}{2(N+k+2)} \, \bigl( k+\tfrac{3}{2}  \bigr) \ , 
\end{eqnarray}
since $k^+=(k+1)$ and $k^++k^-=k+N+2$. Thus it follows that $(0;{\rm f})$ saturates precisely the BPS bound. 
Obviously, the same argument also 
applies to $(0;\bar{\rm f})$, for which $l^+=0$, $l^-=\tfrac{1}{2}$,  $\hat{u}=-\tfrac{N+2}{2}$. 
\smallskip

The other simple representation is the $({\rm f};0)$ representation, for which we look for singlets
with respect to $\mathfrak{su}(N)$ in the affine  $\mathfrak{su}(N+2)$ representation based on the 
fundamental representation. The relevant states are simply those states 
from the ground states in the fundamental representation of $\mathfrak{su}(N+2)$ that transform as a singlet with respect
to $\mathfrak{su}(N)$, where the decomposition with respect to $\mathfrak{su}(N)\oplus \mathfrak{su}(2) \oplus \mathfrak{u}(1)$ 
is 
\begin{equation}\label{fdec}
[{\bf N+2}]= ({\bf N},{\bf 1})_{1} + ({\bf 1},{\bf 2})_{-N/2} \ .
\end{equation}
(Here the index denotes the eigenvalue with respect to $\hat{U}$.) The relevant 
states carry therefore the quantum numbers $l^+=\tfrac{1}{2}$, $l^-=0$ and $\hat{u}=-\tfrac{N}{2}$. 
The conformal weight equals
\begin{equation}\label{f0}
h({\rm f};0) = \frac{C^{(N+2)}({\rm f})}{k+N+2} - \frac{N^2}{4N (N+2) (N+k+2)}= \frac{N+\tfrac{3}{2}}{2 (k+N+2)} \ .
\end{equation}
This now has to be compared to the BPS bound which equals in this case
\begin{eqnarray}
h(l^+=\tfrac{1}{2},l^-=0,u=0)_{\rm BPS} & = & \frac{1}{(N+k+2)} \Bigl( \tfrac{1}{2}(N+1) +  \tfrac{1}{4} \Bigr) \\
& = &  \frac{1}{2(N+k+2)} \bigl( N+ \tfrac{3}{2} \bigr) \ . 
\end{eqnarray}
Thus these states saturate also the BPS bound. Note that these two representations are also annihilated by an 
additional supersymmetry generator. From the point of view of representation theory the consideration is identical 
to that in Section~\ref{sec:fund}, see eq.~(\ref{Gshort}).
The generic BPS representation to be considered in the next subsection will only be annihilated by a single generator $G^{++}_{-1/2}$. 

\smallskip

We should mention in passing that if we had not divided out by the $\mathfrak{u}(1)$ current as described in 
Section~\ref{sec:uone}, the two representations above would still have been BPS, but their conformal weight
would have been instead 
\be
h'(0;{\rm f}) = \frac{k+2+\frac{1}{N}}{2(N+k+2)} \ ,   \qquad \quad 
h'({\rm f};0) =  \frac{N+2-\frac{1}{N+2}}{2(N+k+2)} \ ,  
\ee
and their $\mathfrak{u}(1)$ charges would have been $\hat{u}(0;{\rm f}) = \tfrac{N+2}{2}$ and 
$\hat{u}({\rm f};0) = -\tfrac{N}{2}$, respectively. In particular, these quantum numbers do not 
respect the $N\leftrightarrow k$ symmetry under which these two representations should be related
to one another. On the other hand, this symmetry is respected by the results above, see eqs.~(\ref{0f}) and 
(\ref{f0}). 
\medskip

\subsection{Higher Representations}

Next we want to consider the representations that appear in the various products of the above minimal representations.
For example, the representation $({\rm f};{\rm f})$ arises as above from the ground states in the fundamental representation
of $\mathfrak{su}(N+2)$ that transform in the fundamental representation w.r.t.\ $\mathfrak{su}(N)$, i.e.\ from the 
first term in (\ref{fdec}). Together with the fact that $\hat{u} = 1$ we then find
\be\label{ff}
h({\rm f};{\rm f}) = \frac{(N+1)^2}{N(N+2)(N+k+2)} - \frac{1}{N(N+2)(N+k+2)} = \frac{1}{(N+k+2)} \ . 
\ee
This representation does not saturate the BPS bound since it has $l^\pm = 0$ (and $u=0$), and thus the BPS bound is 
simply $h_{\rm BPS}=0$. Note that (\ref{ff}) behaves again as a `light' state, i.e.\ its conformal dimension
vanishes in the 't~Hooft limit.
\medskip

On the other hand, the representation $({\rm f};\bar{\rm f})$ is BPS. Indeed, it arises from the second term
in (\ref{fdec}) upon applying a fermionic generator $\bar\psi^{i,\alpha}$. Its $\hat{U}$-charge is therefore
$\hat{u}= - \tfrac{N}{2} - \tfrac{(N+2)}{2} = - N - 1$, and thus the conformal dimension equals
\be\label{marg}
h({\rm f};\bar{\rm f}\,) = \frac{1}{2} \ . 
\ee
This saturates the BPS bound (\ref{A28}) since the above state has $l^+=l^-=\tfrac{1}{2}$ (as well as $u=0$). 
In fact, it defines a marginal operator by which the conformal field theory may be deformed (without 
destroying the large ${\cal N}=4$ symmetry). 
\medskip

Similarly, we can consider the representations that appear in the products of $({\rm f};0)$ with itself. 
The relevant analysis is done in appendix~\ref{app:reps}, and we only summarise the salient points here.
The fusion rules predict that the two-fold product is of the form
\be
({\rm f};0) \otimes ({\rm f};0) = ([2,0,\ldots,0];0) \oplus ([0,1,0,\ldots,0];0) \ , 
\ee
where the first term corresponds to the symmetric product, while the second term is the `anti-symmetric'
combination. It turns out that the symmetric product is BPS, while the anti-symmetric is not, see 
eqs.~(\ref{4.44}) -- (\ref{4.47}). On the other hand, for the representations of the form $(0;{\rm f})$, the situation 
is reversed in that $(0;[0,1,0,\ldots,0])$ is BPS, while $(0;[2,0,0,\ldots,0])$ is not, compare eqs.~(\ref{B8})
-- (\ref{B11}). However, in either case, it is the representation with $l^\pm=1$ that is BPS.

\subsection{Summary of BPS spectrum}\label{sec:BPSgen}

The above findings suggest that the states $({\rm f};0)$ and $(0;\bar{\rm f})$ perserve the same supercharges,
and therefore that their product is also BPS. Furthermore, among the `multi-particle' states of 
$({\rm f};0)$ or $(0;{\rm f})$, the BPS state is the totally symmetric (or totally anti-symmetric) state, see
appendix~\ref{app:Bgen}. The relevant state always has the maximal spin with
respect to the relevant $\mathfrak{su}(2)$ algebra, e.g.\ in the example of the previous subsection,  we have
$l^+=1$ and $l^-=1$, respectively, see eq.~(\ref{4.45}) and (\ref{B10}).

Extrapolating from the above findings we therefore conclude that the coset theory has BPS states with 
\be\label{BPS1}
l^+ \in \tfrac{1}{2} \mathbb{N}_0 \ ,\qquad  l^- \in \tfrac{1}{2} \mathbb{N}_0 \ , 
\ee
where we have again set the $\mathfrak{u}(1)$ charge to zero. These 
states are the ones that appear in suitable (symmetrised) powers of $({\rm f};0)$ and $(0;\bar{\rm f})$. Obviously,
there is also the charge-conjugate set that is generated by $(\bar{\rm f};0)$ and $(0;{\rm f})$, for which we get
the same quantum numbers 
\be\label{BPS2}
l^+ \in \tfrac{1}{2} \mathbb{N}_0 \ ,\qquad  l^- \in \tfrac{1}{2} \mathbb{N}_0 \ .
\ee
With respect to the $A_\gamma$ algebra these states carry the {\em same}
quantum numbers, but they will differ with respect to the full $s{\cal W}^{(4)}_{\infty}[\gamma]$ algebra, in particular, they will have
opposite eigenvalues for the spin $3$ generator, etc. (The situation is therefore analogous to what happened
in the bosonic case, where $({\rm f};0)$ and $(\bar{\rm f};0)$ define the same Virasoro representation, but have
opposite $W^3_0$ eigenvalue.) The only exception is the state with 
$l^\pm=0$, $u=0$ that is common to both families, and that just defines the vacuum representation. In any case,
the conformal dimensions of all of these representations have the form
\be\label{BPSh}
h  = \frac{1}{N+k+2} \Bigl[ (k+1) l^- + (N+1) l^+ +(l^+-l^-)^2 \Bigr] \ .
\ee
Note that this bound is also identical for the non-linear $\tilde{A}_\gamma$ algebra, see 
eq.~(\ref{BPSnon}), since $\hat{k}^+=k$ and $\hat{k}^-=N$.

\section{Comparison of the Spectrum}\label{sec: compar}

In this section we will make a preliminary comparison of the spectrum of states of the 
${\rm shs}_2[\mu]$ Vasiliev theory of Section~\ref{sec:hspin}, with the dimensions of operators in the coset theory described in 
Section~\ref{sec:coset} and \ref{sec:spec}. Since the Vasiliev description is classical, we can only meaningfully 
compare with the spectrum of the coset theory in the large $N$ ('t~Hooft like) limit.  
As in \cite{Gaberdiel:2010pz}, we define the 't~Hooft limit of the coset by
taking the rank $N$ and level $k$ to infinity, while keeping the combination 
$\frac{N}{N+k}$ fixed. Actually, in our case it is a bit more natural to define 
the 't~Hooft coupling constant $\lambda$ to equal
\be\label{thftdef}
\lambda= \frac{N+1}{N+k+2} = \frac{k^-}{k^++k^-} =\gamma\ .
\ee
Here $\gamma$ is the parameter characterising the large ${\cal N}=4$ algebra as defined in (\ref{A9}). 

\subsection{Symmetry Currents}

We have already seen in Sections~\ref{sec:hscur} and \ref{sec:nlin} that the spectrum of spin currents of 
the truncated coset algebra $s\tilde{\cal W}^{(4)}_{\infty}[\gamma]$ matches precisely with that of the
asymptotic symmetry algebra of the higher spin theory of Section~\ref{sec:hspin}. In particular, this implies
that the one loop determinants for the higher spin gauge fields computed using the results of \cite{David:2009xg} 
will match the vacuum character of the coset theory. This matching is a straightforward extension of the result
of \cite{Gaberdiel:2010ar} for the bosonic case, and of \cite{Creutzig:2011fe} for the supersymmetric case.

Actually, we can be more specific about the relation between the two parameters since both algebras
contain   the global symmetry algebra $D(2,1|\alpha)$ as a subalgebra. In the Vasiliev theory, 
the parameter $\alpha$ is related to the parameter $\mu$ characterising the shs$_2[\mu]$ higher spin algebra 
by the relation (\ref{parrel}), i.e.\ $\alpha = \frac{\mu}{1-\mu}$. 
On the other hand, from the coset point of view, we saw that the relation is 
$\alpha =\frac{\gamma}{1-\gamma}$, see eq.~(\ref{alphadef}). In other words, 
for the symmetry algebras to be the same on both sides we need to identify the parameters 
$\mu=\gamma$. From (\ref{thftdef}) we see that this implies $\mu=\lambda$.   
We will soon see an independent verification of this identification.

\subsection{Nontrivial Primaries}

We saw in Section~\ref{sec:minrep} that the minimal representations of the coset, labelled as 
$(0; {\rm f})$ and $({\rm f};0)$ (together with their complex conjugates), have their lowest spin zero 
components transforming as $({\bf 1}, {\bf 2})$  and $({\bf 2}, {\bf 1})$ under 
$\mathfrak{su}(2)\oplus \mathfrak{su}(2)$, respectively. We see then that this matches with the quantum numbers of the 
basic scalar fields in the ${\rm shs}_2[\mu]$ higher spin theory. The single particle excitations of the latter are the lowest 
components of the minimal representations of ${\rm shs}_2[\mu]$ whose quantum numbers are given in (\ref{matqno}) 
and are exactly those of the coset. So we are led to the correspondence 
\be
({\rm f};0) \leftrightarrow \phi_+ \ , \qquad\qquad (0; \bar{\rm f}) \leftrightarrow \phi_-  \ .
\ee
We also know that the mass $M_{\pm}$ of $\phi_{\pm}$ are given as in (\ref{scalmass}). 
This corresponds to conformal dimensions for the corresponding primary operators in the CFT to equal
$h_+ = \frac{\mu}{2}$ and $h_- = \frac{1-\mu}{2}$, respectively. We can now compare this to the exact expressions 
for the coset representations in 
(\ref{f0}) and (\ref{0f}). We find that in the 't~Hooft limit, using the definition (\ref{thftdef})
\be
h({\rm f};0) = \frac{N+\frac{3}{2}}{2 (k+N+2)} \ \longrightarrow \ \frac{\lambda}{2} 
\ee
and 
\be
h(0;\bar{\rm f}) = \frac{k+\frac{3}{2}}{2 (k+N+2)} \ \longrightarrow \ \frac{1-\lambda}{2}\ .
\ee
Thus the conformal dimensions also match in the large $N$ 't~Hooft limit, if we make the identification between 
the higher spin algebra parameter $\mu$ and the `t~Hooft parameter $\lambda$ as $\mu=\lambda$. This reproduces
what was found at the end of the previous subsection, and thus furnishes an independent check of the correspondence.

We can go further and compare the BPS spectrum of Section~\ref{sec:BPSgen}. The spectrum of the lowest
scalar components in (\ref{BPSh})  becomes in the large $N$ limit
\be
h(l^+, l^-) \ \rightarrow \ (2l^+)\, \frac{\lambda}{2} + (2l^-)\, \frac{1-\lambda}{2}\ .
\ee
Thus we have a tower of states labelled by the two non-negative integers $2l^{\pm} =0, 1, 2\ldots $. This precisely 
corresponds to the spectrum of multi-particle states with $2l^+$ excitations of $\phi_+$, and $2l^-$ excitations of 
$\phi_-$ in the classical Vasiliev theory. The energies are simply additive since the bulk theory is free 
$(G_N \propto \frac{1}{c} \sim \frac{1}{N})$. This provides further non-trivial evidence for the claim that the 
large ${\cal N}=4$ coset theory (\ref{N4coset2}) in the large $N$ 't~Hooft limit is 
captured by a classical Vasiliev theory on AdS$_3$ based on the shs$_2[\mu]$ higher spin algebra. 

On the other hand, the 't~Hooft limit of the coset also contains a spectrum of `light states'. In particular, the
conformal dimension of $({\rm f};{\rm f})$ equals (see eq.~(\ref{ff}))
\be
h({\rm f};{\rm f}) = \frac{1}{N+k+2} \ \longrightarrow \ \frac{\lambda}{N}
\ee
in the 't~Hooft limit. In fact, there will be a continuum of such states corresponding to the representations of the
form $(\Lambda;\Lambda)$ since we have 
\begin{eqnarray}
h(\Lambda;\Lambda) & = &  \frac{C^{(N+2)}(\Lambda)}{k+N+2} - \frac{C^{(N)}(\Lambda)}{k+N+2} - 
\frac{|\Lambda|^2}{N (N+2) (N+k+2)} \\
& \cong & \frac{|\Lambda|}{k+N+2} \, \bigl( 1 - \tfrac{|\Lambda|}{N(N+2)} \bigr) \ \longrightarrow \ \frac{ \lambda |\Lambda|}{N} \ ,
\end{eqnarray}
where $|\Lambda|$ denotes the number of boxes (and anti-boxes) of $\Lambda$. Nevertheless, we expect the
large $N$ 't~Hooft limit to be sensible, compare the discussion in 
\cite{Papadodimas:2011pf,Chang:2011vk,Jevicki:2013kma,Chang:2013izp}.

We take note of a special operator in the BPS spectrum, namely, the primary labelled 
$({\rm f};\bar{\rm f})$ which has $h({\rm f};\bar{\rm f}\,) = \tfrac{1}{2}$. Just as in the case of 
the ${\cal N}=2$ superconformal algebra, such a chiral operator has a descendant with $h=1$  
which (together with its right moving partner) is a marginal supersymmetry preserving operator. Turning on this operator thus preserves the large ${\cal N}=4$ superconformal algebra but would 
generically break the higher spin symmetries of the coset. This is natural from the bulk point of view since this operator is a double trace operator formed from the single trace 
$({\rm f};0)$ and $(0;\bar{\rm f})$ operators. Thus from the bulk point of view it corresponds to changing the boundary conditions of the scalar/fermion field. One expects that this will break the higher symmetry along the lines described in a similar case in \cite{Chang:2013izp}.

\section{Relation to String Theory on ${\rm AdS}_3\times {\rm S}^3\times {\rm S}^3\times {\rm S}^1$}\label{sec:stringy}

As mentioned in the introduction, there is a natural type IIB string theory background with  large ${\cal N}=4$ supersymmetry 
\cite{Boonstra:1998yu,Elitzur:1998mm, de Boer:1999rh, Gukov:2004ym}. The background geometry is 
${\rm AdS}_3\times {\rm S}^3\times {\rm S}^3\times {\rm S}^1$ with 3-form fluxes on both ${\rm S}^3$'s (as well as the 
${\rm AdS}_3$). The background is characterised by three integers, conventionally denoted by the two D5-brane charges 
$Q_5^{\pm}$ and a D-string charge $Q_1$. The Brown-Henneaux central charge of the CFT$_2$ dual to this AdS$_3$ 
background is given by
\be\label{stringc}
c =\frac{6\, Q_1Q_5^+Q_5^-}{Q_5^++Q_5^-}\ .  
\ee 
This is of the general (linear) $A_{\gamma}$ form of the central charge as given in (\ref{A9}) with the two 
$\mathfrak{su}(2)$ levels being equal to  $k^{\pm}=Q_1Q_5^{\pm}$. 

An analysis of the supergravity spectrum \cite{de Boer:1999rh} gives a BPS spectrum of $D(2,1|\alpha)$ which may plausibly be 
organised into BPS multiplets of the linear $A_{\gamma}$ algebra \cite{Gukov:2004ym}\footnote{The caveat is due to 
multiplets saturating the $D(2,1|\alpha)$  BPS bound (\ref{A30}) not obviously saturating the $A_{\gamma}$ bound 
(\ref{A28}) for $l^+\neq l^-$. They saturate the $A_{\gamma}$ bound only if their  masses (dimensions) get appropriate 
quantum $\frac{1}{k}$ corrections.}. The result is an $A_{\gamma}$ BPS spectrum labelled by $(l^+, l^-)$, where 
$2l^{\pm}$ are non-negative integers denoting the $\mathfrak{su}(2)_{k^{\pm}}$ quantum 
numbers\footnote{The $\mathfrak{su}(2)$ quantum numbers on left and right are the same.} of the two 
${\rm S}^3$'s. Each such multiplet comes with multiplicity one, with 
$(l^+=0, l^-=0)$ being the vacuum representation. Proposed duals involving the symmetric product of 
$({\rm S}^3\times {\rm S}^1)$ only possess short $A_{\gamma}$ multiplets with $l_+=l_-$.

In our coset family we have seen that we have a whole tower of $A_{\gamma}$ BPS states (on the left as well as the right) which have arbitrary $(l_+, l_-)$ with multiplicity one. In the large $N$ 't~Hooft limit we interpreted these in the bulk as 
multi-particle states (multiplets) built from the scalars corresponding to the representations $\phi_{\pm}$. Thus we do seem to easily get a tower of states with the right quantum numbers. However, they are mostly multi-particle states. 

But this immediately suggests how we can get a tower of {\it single} particle states in the bulk with arbitrary  $(l_+, l_-)$ and multiplicity one. We simply promote the bulk scalars/fermions to non-abelian $M\times M$ valued fields, and restrict ourselves to ${\rm U}(M)$ singlets. We then take the same suitably symmetrised/antisymmetrised powers of the scalars which was a BPS configuration and take its trace. This is now a single particle state from the point of view of the bulk. For sufficiently large $M$, we will therefore get a tower of single particle states with arbitrary $(l_+, l_-)$. They will appear with multiplicity one for the same reason that it was only a certain symmetrised combination of the bulk scalars which was BPS. 

In particular, there is exactly one (complex) BPS state with $(l^+=\tfrac{1}{2}, l^- =\tfrac{1}{2})$ i.e.\ with $h=\bar{h}= \tfrac{1}{2}$. This multiplet has a descendant state with $h=\bar{h}=1$ which corresponds to a marginal operator that preserves the large ${\cal N}=4$ SUSY. This is exactly what one sees in the string theory as well where there is exactly one complex modulus (see \cite{Gukov:2004ym} for a full discussion). It will be interesting to try and match the detailed properties of this modulus with that seen by the non-abelian Vasiliev theory. As mentioned earlier, turning on this operator very likely breaks the higher spin symmetry. This is as one might expect when going away from the `tensionless' limit which is at the opposite extreme to the supergravity limit in the moduli space of the string theory. 

There is one subtle point we should mention: since the background geometry involves an ${\rm S}^1$ factor, the dual 
CFT should contain a $\mathfrak{u}(1)$ current algebra, and hence really involve the linear $A_\gamma$ algebra (rather 
than $\tilde{A}_\gamma$). On the other hand, from the point of view of the higher spin theories, we seem to get
the non-linear $\tilde{A}_\gamma$ algebra, rather than $A_\gamma$. However, it seems plausible that one can add
the corresponding degrees of freedom, i.e.\ $4$ free fermions and a ${\rm U}(1)$ gauge field, to the non-abelian Vasiliev 
theory so that the asymptotic symmetry algebra contains the linear $A_\gamma$ algebra.

Assuming that this can be done, it seems that a non-abelian version of the Vasiliev theory we have constructed in the bulk 
has the right BPS spectrum to correspond to string theory on 
${\rm AdS}_3\times {\rm S}^3\times {\rm S}^3\times {\rm S}^1$. Note that for consistency of the higher spin symmetry we need to have all the higher spin fields take values in the adjoint of ${\rm U}(M)$ as well. However, once again we restrict ourselves to singlet states. This is equivalent to saying that we gauge the global ${\rm U}(M)$ symmetry on the boundary and thus consider only singlet states under 
${\rm U}(M)$ in the boundary CFT. For large $M$ we might view this phenomenon as a dynamic confinement in the bulk of ${\rm U}(M)$ 
since the bulk 't~Hooft coupling $g_B^2 \propto \frac{M}{N} \approx {\cal O}(1)$, as observed in \cite{Chang:2012kt}. 

While the ${\rm U}(M)$ Vasiliev theory at large $M$ seems to be on the right track, the obvious coset candidates, e.g.\ the cosets
\be
\frac{\mathfrak{su}(N+2M)^{(1)}_\kappa}{\mathfrak{su}(N)^{(1)}_\kappa \oplus \mathfrak{su}(M)^{(1)}_{\kappa}}
\ee
corresponding to a ${\rm U}(M)$ gauging, do not appear to work. It would be very interesting to identify the
coset constructions that are dual to the ${\rm U}(M)$ singlet sector of the non-abelian higher spin theory.

\section{Conclusions}\label{sec:concl}

In this paper we have constructed a higher spin theory based on the higher spin algebra shs$_2[\mu]$, which contains
in particular the exceptional superalgebra $D(2,1|\alpha)$ as a subalgebra. The higher spin theory therefore preserves
the large ${\cal N}=4$ supersymmetry. We have also identified a candidate dual 2d CFT: it is given by the 
't~Hooft limit of the large ${\cal N}=4$ cosets corresponding to the Wolf symmetric spaces. We have shown that the asymptotic
symmetry algebra of the higher spin theory matches the $s\tilde{\cal W}^{(4)}_{\infty}[\gamma]$ algebra of the
(truncated) cosets in the 't~Hooft limit. Since both contain $D(2,1|\alpha)$ as a subalgebra, we could identify
the $\mu$ parameter of the higher spin theory with the usual 't~Hooft parameter $\lambda$ of the large $N$ limit. This 
identification was then subsequently confirmed by comparing the BPS spectra of the two descriptions.

There is a natural string solution with large ${\cal N}=4$ supersymmetry, whose background geometry is 
AdS$_3\times {\rm S}^3 \times {\rm S}^3 \times {\rm S}^1$. We have argued that the  corresponding supergravity
spectrum can be accounted for in terms of a non-abelian generalisation of the above Vasiliev theory, in close
analogy to what was proposed in one dimension higher in \cite{Chang:2012kt}. This opens the 
exciting possibility of understanding the relation between higher spin theory and string theory for this 
very controlled setting in detail. One may also hope to use the insights from the higher spin description in order
to find the CFT dual to the AdS$_3\times {\rm S}^3 \times {\rm S}^3 \times {\rm S}^1$ string. 

Another interesting direction to study are the cases with ${\cal N}=2$ supersymmetry. In particular, the analogues
of the non-abelian generalisation of the ${\cal N}=2$ higher spin theories are quite plausibly related to the
general Kazama-Suzuki models corresponding to 
\be\label{cosets1}
\frac{\mathfrak{su}(N+M)^{(1)}_{\kappa}}{\mathfrak{su}(N)^{(1)}_{\kappa} \oplus
\mathfrak{su}(M)^{(1)}_{\kappa} \oplus \mathfrak{u}^{(1)}}  \ ,
\ee
where $\kappa=k+N+M$. (Indeed, the cosets  with $M=1$ describe the CFT duals of the ${\cal N}=2$ higher
spin theory \cite{Creutzig:2011fe}, and it seems plausible that the cosets with $M>1$ correspond to the non-abelian
generalisation of the ${\cal N}=2$ higher spin theory.) 
In the `stringy' limit in which $M$, $N$, and $k$ are simultaneously taken
to infinity, the central charge 
\be\label{kscent}
c = \frac{3 k M N }{M+N+k} 
\ee
is proportional to $N^2$, as appropriate for a stringy model. Furthermore, the light states that appear in the 't~Hooft limit
$N,k\rightarrow \infty$ for fixed $M$ become lifted in the limit where all three quantum numbers become large
simultaneously. It would be very interesting to identify the dual string backgrounds that may correspond to these
interpolating coset theories. Another example of an ${\cal N}=2$ `stringy coset' where it would be very interesting to 
understand the dual string background is the one studied in \cite{Gopakumar:2012gd}.

\section*{Acknowledgements}

We thank Misha Vasiliev for a useful discussion about the structure of 
${\rm shs}_M[\mu]$ and Shiraz Minwalla for discussions on string theories on AdS$_3$.  
We would also like to thank the participants of the GGI Conference on `Higher Spins, Dualities and Strings' 
for their feedback. 
The work of MRG is supported in parts by the Swiss National
Science Foundation. RG's research is partially supported by a 
Swarnajayanthi fellowship of the DST, Govt.\ of India, and more broadly by the 
generosity of the Indian people towards basic sciences. Finally, we thank the Galileo Galilei Institute for 
Theoretical Physics for the hospitality and the INFN for partial support during the final stages of this work.

\appendix

\section{The Global Superalgebra  $D(2,1|\alpha)$}\label{app:D21}

The global symmetry algebra that is relevant in our context is the exceptional superalgebra
$D(2,1|\alpha)$ that is generated by 
\be\label{D21alpha}
L_0\ , L_{\pm 1} \ , \quad G^a_{\pm \frac{1}{2}} \ , \qquad A^{\pm, i}_{0} \ . 
\ee
Here $a\in\{0,1,2,3\}$ and $i\in\{1,2,3\}$, and the commutation relations are 
\begin{eqnarray}
{}[L_m,L_n] & = & (m-n)\, L_{m+n} \\
{}[L_m, G^a_{r}] & = & (\tfrac{m}{2} - r) G^a_{m+r} \\
{}[A^{\pm, i}_{0}, G^a_r] & = & {\rm i}\, \alpha^{\pm\, i}_{ab} G^b_r  \label{B6}\\
{}[A^{\pm , i}_0,A^{\pm, j}_0] & = & {\rm i} \, \epsilon^{ijl} A^{\pm, l}_0 \\
{}\{G^a_r, G^b_s\} & = & 2 \delta^{ab} \, L_{r+s}
+ 4\, (r-s)\, \left(\gamma \,  {\rm i}\, \alpha^{+\, i}_{ab}\, A^{+, i}_{r+s} + (1-\gamma) \, {\rm i}\, \alpha^{-\, i}_{ab}\, A^{-, i}_{r+s} \right) \ ,
\end{eqnarray}
while $[L_m,A^{\pm, i}_0]=0$. Furthermore, the expressions $\alpha^{\pm\, i}_{ab}$ are the $4\times 4$ matrices
\begin{equation}
\alpha^{\pm\, i}_{ab} = \frac{1}{2} \Bigl( \pm \delta_{ia} \delta_{b0} \mp \delta_{ib} \delta_{a0} + \epsilon_{iab} \Bigr) \ ,
\end{equation}
that satisfy the relations 
\begin{equation}
{}[\alpha^{\pm\, i},\alpha^{\pm\, j}] = -\epsilon^{ijl}\, \alpha^{\pm\, l} \ , \qquad
{}[\alpha^{+\, i},\alpha^{-\, j}] = 0 \ , \qquad
{}\{\alpha^{\pm\, i},\alpha^{\pm\, j} \} = - \tfrac{1}{2}\, \delta^{ij} \ .
\end{equation}
Finally, the parameter $\alpha$ in $D(2,1|\alpha)$ equals
\be\label{alphgam}
\alpha = \frac{\gamma}{1-\gamma} \ . 
\ee
Note that the algebra is isomorphic under $\gamma\leftrightarrow (1-\gamma)$; in terms of $\alpha$ this
is the transformation $\alpha\leftrightarrow \alpha^{-1}$. 

\subsection{A Complex Basis}\label{app:comp}

It is sometimes convenient to work with a complex basis where we introduce the Cartan-Weyl generators
for the two $\mathfrak{su}(2)$ algebras, i.e.\ the generators
\be
A^{\pm \, \alpha}_{0} \ , \qquad \alpha \in \{\pm, 3\} 
\ee
with commutation relations of the form
\begin{equation}
{}[A^{*\, 3}_{0},A^{*\,\pm}_{0}]  =  \pm \, A^{*\, \pm}_{0}  \ , \qquad
[A^{*\, +}_{0},A^{*\, -}_{0} ] =  2 \, A^{*\, 3}_{0} \ , 
\end{equation}
where $*$ is either $*=+$ or $*=-$. We can similarly introduce a complex basis for the supercharges via
\be\label{Gcomp}
G^{++}_r = - (G^1_r+{\rm i} \,G^2_r) \ , \ \
G^{+-}_r =  (G^3_r+{\rm i} \,G^0_r) \ , \ \ 
G^{-+}_r =  (G^3_r-{\rm i} \,G^0_r) \ , \ \
G^{--}_r =  (G^1_r-{\rm i} \,G^2_r) \ ,
\ee
and then the commutation relations (\ref{B6}) become
\be\label{A.17a}
\begin{array}{lcllcl}
{}[A^{+\, 3}_{0}, G^{\pm *}_{r}] & =& \pm \tfrac{1}{2} \, G^{\pm *}_{r}  \qquad & 
{}[A^{+\, +}_{0}, G^{+ *}_{r}] & =& 0 \\
{}[A^{+\, -}_{0}, G^{+ *}_{r}] & =& G^{- *}_{r} \qquad & 
{}[A^{+\, +}_{0}, G^{- *}_{r}] & =& G^{+ *}_{r} \\
{}[A^{+\, -}_{0}, G^{- *}_{r}] & =& 0 \ , & & 
\end{array}
\ee
where again $*$ is either $*=+$ or $*=-$. Similarly, the commutationr relations with the other $\mathfrak{su}(2)$
currents take the form
\be\label{A.30a}
\begin{array}{lcllcl}
{}[A^{-\, 3}_{0}, G^{*\pm }_{r}] & =& \pm \tfrac{1}{2} G^{*\pm }_{r} \qquad &
{}[A^{-\, +}_{0}, G^{*+}_{r}] & =& 0 \\
{}[A^{-\, -}_{0}, G^{*+}_{r}] & =& G^{*-}_{r} \qquad & 
{}[A^{-\, +}_{0}, G^{*-}_{r}] & =& G^{*+}_{r} \\
{}[A^{-\, -}_{0}, G^{*-}_{r}] & =& 0 \ . & & 
\end{array}
\ee
Finally, the anti-commutators of the supercharges are then
\be\label{GGpm}
\begin{array}{rcl}
{}\{G^{++}_r, G^{++}_s \} & = & 0 \\
{}\{G^{++}_r, G^{+-}_s \} & = & 4 (r-s) \,\gamma\,  A^{++}_{r+s} \\
{}\{G^{++}_r, G^{-+}_s \} & = & 4 (r-s) \, (1-\gamma) \, A^{-+}_{r+s} \\
{}\{G^{++}_r, G^{--}_s \} & = & - 4 L_{r+s}  
- 4 (r-s) \bigl[ \gamma A^{+3}_{r+s} + (1-\gamma) A^{-3}_{r+s} \bigr] \\
{}\{G^{+-}_r, G^{+-}_s \} & = & 0 \\
{}\{G^{+-}_r, G^{-+}_s \} & = &  4 L_{r+s}  
+ 4 (r-s) \bigl[ \gamma A^{+3}_{r+s} - (1-\gamma) A^{-3}_{r+s} \bigr]  \\
{}\{G^{+-}_r, G^{--}_s \} & = & - 4 (r-s) \, (1-\gamma) \, A^{--}_{r+s} \\
{}\{G^{-+}_r, G^{-+}_s \} & = & 0 \\
{}\{G^{-+}_r, G^{--}_s \} & = &  - 4 (r-s) \, \gamma \, A^{+-}_{r+s} \\\
{}\{G^{--}_r, G^{--}_s \} & = & 0 \ .
\end{array}
\ee

\subsection{BPS Representations of $D(2,1|\alpha)$}

The highest weight representations of $D(2,1|\alpha)$ are labelled by $l^+, l^-, h$, where $l^\pm$ is
the spin of the two $\mathfrak{su}(2)$ algebras generated by $A^{\pm\, i}_0$, while $h$ is the eigenvalue of $L_0$. 
(The highest weight states are annihilated by the positive modes, $G^a_{1/2}$ and $L_1$.)
There is a unitarity bound that arises from requiring the norm of 
\begin{equation}\label{A29}
{\cal N}_2^{\prime} = G^{--}_{-1/2} \, | (h,l^\pm)\rangle \ ,
\end{equation}
to be positive (where $(G^{--}_{-1/2})^\dagger = -G^{++}_{1/2}$); it takes the form
\begin{equation}\label{A30}
h \geq \Bigl[ \frac{1}{1+\alpha}\, l^- + \frac{\alpha}{1+\alpha} \, l^+ \Bigr]  \ .
\end{equation}

\medskip

For the truncation analysis of Section~\ref{sec:asymp} another
class of short representations plays an important role, namely the representations of the
form $\hat{R}^{(N)}_{\pm}$, see (\ref{D2multshort}). Let $\alpha=-\tfrac{s+1}{s}$, i.e.\
$\gamma=\mu=s+1$. Then the representation $\hat{R}^{(s)}_{-}$ is generated from
a state $\Phi_s$, satisfying 
\be
L_0 \Phi_s = s\, \Phi_s \ , \qquad L_1 \Phi_s = 0 \ , \qquad
G^{\alpha\beta}_{\frac{1}{2}} \Phi_s = 0 \ ,
\ee
with $\Phi_s$ being a singlet with respect to the two $\mathfrak{su}(2)$ algebras. This representation
then contains an ideal that is generated by the state
\be\label{idealgen}
{\cal N} = G^{+-}_{-\frac{1}{2}} \, G^{++}_{-\frac{1}{2}} \, \Phi_s \ .
\ee
This state transforms in the $({\bf 3},{\bf 1})$ with respect to the two $\mathfrak{su}(2)$ algebras, and thus
quotienting out this ideal leads to the spectrum of $\hat{R}^{(s)}_{-}$. In order to show that it actually defines an
ideal one calculates
\begin{eqnarray}
G^{--}_{1/2} \, {\cal N} & = & 4 (1-\gamma) A^{--}_0  G^{++}_{-1/2} \, \Phi_s 
- G^{+-}_{-1/2} \bigl( - 4 L_0 + 4 (\gamma A^{+3}_{0} + (1-\gamma) A^{-3}_0 ) \bigr) \Phi_s \nonumber \\
& = & \bigl( 4 (1-\gamma)   + 4 s  \bigr)\, G^{+-}_{-1/2} \, \Phi_s = 0  \ , 
\end{eqnarray}
where we have first used that $A^{\pm a}_0 \Phi_s=0$ since $\Phi_s$ is a singlet, and then
$\gamma=s+1$. We should mention that short representations of this kind are rather unusual, since
the ideal only appears at the `second level', and is not directly visible on the ground states; in particular
$\Phi_s$ does not saturate the BPS bound (\ref{A30}) since $l^\pm=0$ and $h=s$.

\section{The Large ${\cal N}=4$ Algebra}\label{app:alg}

Next let us review the structure of the large ${\cal N}=4$ algebra. We begin with the linear $A_\gamma$ algebra, for which
the various non-trivial (anti-)commutators are \cite[eq.~(4.3)]{Gukov:2004ym}\footnote{Relative to the conventions of 
\cite{Gukov:2004ym} we have rescaled the currents $U$ and $A^{\pm\, i}$, as well as the
$Q^a$ fields by a factor of ${\rm i}$, in order to remove some minus signs.}
\begin{eqnarray}
{}[U_m,U_n] & = &  \tfrac{k^+ + k^-}{2} \, m \, \delta_{m,-n}  \label{A1} \\[2pt]
{}[A^{\pm, i}_m, Q^a_r] & = & {\rm i}\, \alpha^{\pm\, i}_{ab} \, Q^b_{m+r}   \label{A2} \\[4pt]
{} \{Q^a_r,Q^b_s \} & = &  \tfrac{k^+ + k^-}{2} \, \, \delta^{ab} \, \delta_{r,-s}  \label{A3} \\[2pt]
{}[A^{\pm, i}_{m}, A^{\pm, j}_{n} ] & = &  \tfrac{k^\pm}{2}\, m \, \delta^{ij}\, \delta_{m,-n} 
+ {\rm i}\, \epsilon^{ijl}\, A^{\pm, l}_{m+n}  \label{A4} \\[2pt]
{} [U_m,G^a_r] & = & m \, Q^a_{m+r}  \label{A5} \\[2pt]
{}[A^{\pm, i}_{m},G^a_r] & = &  {\rm i} \,\alpha^{\pm\, i}_{ab} \,G^b_{m+r}  
\mp \tfrac{2 k^\pm }{k^++k^-}\, m \, \alpha^{\pm\, i}_{ab}\, Q^b_{m+r}  \label{A6} \\[2pt]
{} \{Q^a_r,G^b_s\} & = & 2\,  \alpha^{+\, i}_{ab}\, A^{+, i}_{r+s} - 2\,  \alpha^{-\, i}_{ab}\, A^{-, i}_{r+s} + \delta^{ab} \,
U_{r+s}  \label{A7} \\[4pt]
{} \{G^a_r,G^b_s\} & = & \tfrac{c}{3}\, \delta^{ab}\, (r^2 - \tfrac{1}{4}) \delta_{r,-s} 
+ 2\, \delta^{ab}\, L_{r+s} \nonumber \\
& & \ + 4\, (r-s)\, \left(\gamma \,  {\rm i}\, \alpha^{+\, i}_{ab}\, A^{+, i}_{r+s} + (1-\gamma) \, {\rm i}\, \alpha^{-\, i}_{ab}\, A^{-, i}_{r+s} \right) \ ,
 \label{A8}
\end{eqnarray}
where 
\begin{equation}\label{A9}
\gamma = \frac{k^-}{k^+ + k^-} \ , \qquad c = \frac{6 k^+ k^-}{k^+ + k^-} \ . 
\end{equation}
In addition, the commutators with the Virasoro modes $L_m$ (that satisfy the usual Virasoro algebra with 
central charge $c$) take the familiar form, i.e.\ 
\begin{equation}
{}[L_m, V_n] = ((h-1)m - n) V_{m+n} \qquad \hbox{if $V$ has conformal dimension $h$.}
\end{equation}
The conformal dimensions of the fields $Q^a$, $U$, $A^{\pm, i}$, and $G^a$ are $h=\tfrac{1}{2}$, $h=1$, $h=1$, 
and $h=\tfrac{3}{2}$, respectively. The parameters $a,b$ take the values $a,b\in\{0,1,2,3\}$, while
the indices $i,j,l$ are vector indices and take the values $i,j,l\in \{1,2,3\}$.
Note that the large ${\cal N}=4$ algebra $A_\gamma$ contains the current algebras 
\begin{equation}
\mathfrak{su}(2)_{k^+} \oplus \mathfrak{su}(2)_{k^-} \oplus \mathfrak{u}(1) 
\end{equation}
that are generated by the $A^{\pm, i}$ fields, as well as the $U$ field, and that commute with one another. In addition, 
there are $4$ supercharges corresponding to $G^a$ that transform in the $(\tfrac{1}{2},\tfrac{1}{2})_0$ representation
with respect to $\mathfrak{su}(2) \oplus \mathfrak{su}(2)\oplus \mathfrak{u}(1)$.

Note that the `wedge algebra' (where we restrict to the modes $V_n$ with $|n| <h$) is isomorphic to $D(2,1|\alpha)$ 
together with a central element corresponding to $U_0$, where we have the relation
\be\label{alphadef}
\alpha = \frac{\gamma}{1-\gamma} \ .
\ee

\subsection{A Complex Basis}

Again, we can introduce a complex basis for the currents, $A^{\pm\, \alpha}_m$, for which we have the affine 
commutation relations
\begin{eqnarray}
{}[A^{*\, 3}_{m},A^{*\,\pm}_{n}] & = & \pm \, A^{*\, \pm}_{m+n} \\
{}[A^{*\, +}_{m},A^{*\, -}_{n} ]& = & 2 \, A^{*\, 3}_{m+n} + k^* m\, \delta_{m,-n} \\
{}[A^{*\, 3}_{m},A^{*\, 3}_{n} ]& = & \tfrac{k^*}{2} m \, \delta_{m,-n} \ , 
\end{eqnarray}
where $*$ is either $*=+$ or $*=-$. Introducing a complex basis for the supercurrents and the free fermions as in 
(\ref{Gcomp}), the commutation relations (\ref{A6}) become
\begin{eqnarray}
{}[A^{+\, 3}_{m}, G^{\pm *}_{r}] & =& \pm \tfrac{1}{2} \Bigl( G^{\pm *}_{m+r} - \tfrac{2k^+}{k^++k^-}\, m \,Q^{\pm *}_{m+r}\Bigr) 
\label{A.17}
\\[4pt]
{}[A^{+\, +}_{m}, G^{+ *}_{r}] & =& 0 \\
{}[A^{+\, -}_{m}, G^{+ *}_{r}] & =& \Bigl( G^{- *}_{m+r} - \tfrac{2k^+}{k^++k^-}\, m \,Q^{- *}_{m+r}\Bigr) 
\\[4pt]
{}[A^{+\, +}_{m}, G^{- *}_{r}] & =& \Bigl( G^{+ *}_{m+r} - \tfrac{2k^+}{k^++k^-}\, m \,Q^{+ *}_{m+r}\Bigr)  \\[4pt]
{}[A^{+\, -}_{m}, G^{- *}_{r}] & =& 0 \ , 
\end{eqnarray}
where again $*$ is either $*=+$ or $*=-$. Similarly, the commutation relations with the other $\mathfrak{su}(2)$
currents take the form
\begin{eqnarray}
{}[A^{-\, 3}_{m}, G^{*\pm }_{r}] & =& \pm \tfrac{1}{2} \Bigl( G^{*\pm }_{m+r} + \tfrac{2k^-}{k^++k^-}\, m \,Q^{*\pm}_{m+r}\Bigr) 
\\[4pt]
{}[A^{-\, +}_{m}, G^{*+}_{r}] & =& 0 \\
{}[A^{-\, -}_{m}, G^{*+}_{r}] & =& \Bigl( G^{*-}_{m+r} + \tfrac{2k^-}{k^++k^-}\, m \,Q^{*-}_{m+r}\Bigr) 
\\[4pt]
{}[A^{-\, +}_{m}, G^{*-}_{r}] & =& \Bigl( G^{*+}_{m+r} + \tfrac{2k^-}{k^++k^-}\, m \,Q^{*+}_{m+r}\Bigr)  \\[4pt]
{}[A^{-\, -}_{m}, G^{*-}_{r}] & =& 0 \ . \label{A.30}
\end{eqnarray}
Finally, the anti-commutators of the supercharges are then
\be\label{GGpmA}
\begin{array}{rcl}
{}\{G^{++}_r, G^{++}_s \} & = & 0 \\
{}\{G^{++}_r, G^{+-}_s \} & = & 4 (r-s) \,\gamma\,  A^{++}_{r+s} \\
{}\{G^{++}_r, G^{-+}_s \} & = & 4 (r-s) \, (1-\gamma) \, A^{-+}_{r+s} \\
{}\{G^{++}_r, G^{--}_s \} & = & - 4 L_{r+s} - \tfrac{2c}{3} (r^2-\tfrac{1}{4}) \delta_{r,-s} 
- 4 (r-s) \bigl[ \gamma A^{+3}_{r+s} + (1-\gamma) A^{-3}_{r+s} \bigr] \\
{}\{G^{+-}_r, G^{+-}_s \} & = & 0 \\
{}\{G^{+-}_r, G^{-+}_s \} & = &  4 L_{r+s} + \tfrac{2c}{3} (r^2-\tfrac{1}{4}) \delta_{r,-s} 
+ 4 (r-s) \bigl[ \gamma A^{+3}_{r+s} - (1-\gamma) A^{-3}_{r+s} \bigr]  \\
{}\{G^{+-}_r, G^{--}_s \} & = & - 4 (r-s) \, (1-\gamma) \, A^{--}_{r+s} \\
{}\{G^{-+}_r, G^{-+}_s \} & = & 0 \\
{}\{G^{-+}_r, G^{--}_s \} & = &  - 4 (r-s) \, \gamma \, A^{+-}_{r+s} \\\
{}\{G^{--}_r, G^{--}_s \} & = & 0 \ .
\end{array}
\ee

We can also identify an ${\cal N}=2$ superconformal algebra within the large ${\cal N}=4$ algebra,
see also \cite{Sevrin:1988ew,Gunaydin:1988re}. Indeed, we can identify the supercharges of the ${\cal N}=2$
algebra with 
\begin{equation}
G^+ = \frac{{\rm i}}{\sqrt{2}} \,  G^{++} \ , \qquad G^- = \frac{{\rm i}}{\sqrt{2}}\, G^{--} 
\end{equation}
and the ${\rm U}(1)$ current with 
\be
J = 2 \Bigl( \gamma A^{+3} + (1-\gamma) A^{-3} \Bigr) \ . 
\ee
It is easy to see that they then generate the commutation relations of the ${\cal N}=2$ algebra, in particular
\begin{eqnarray}
\{ G^+_{r}, G^{-}_s \} & = & 2 L_{r+s} + (r-s) J_{r+s} + \frac{c}{3} (r^2-\tfrac{1}{4}) \delta_{r,-s} \\
{} [J_m,G^\pm_r] & = & \pm G^\pm_{m+r} \ , \qquad 
[J_m,J_n] = \frac{c}{3} m \, \delta_{m,-n} \ . 
\end{eqnarray}

\subsection{The BPS Bound}\label{app:bound}

The representations of the large ${\cal N}=4$ algebra $A_\gamma$ are characterised by $(h,l^\pm,u)$, where
$h$ is the conformal dimension, $l^\pm$ are the spins of the two affine $\mathfrak{su}(2)$ algebras,
and $u$ denotes the ${\rm U}(1)$-charge, i.e.\ the eigenvalue under $U_0$. If we require unitarity,
we need that  $l^\pm \leq k^\pm/2$.  However, as explained in \cite{Gunaydin:1988re}, unitarity actually requires that 
\begin{equation}\label{A.23}
l^\pm \leq \frac{(k^\pm -1)}{2} \ .
\end{equation}
In order to derive the BPS bound that is analogeous to (\ref{A30}) we consider the state
\begin{equation}\label{A26}
{\cal N}_2 =  \Bigl(G^{--}_{-1/2} -  \frac{ 2 (u+{\rm i}(l^+-l^-))}{k^+ + k^-} Q^{--}_{-1/2} \Bigr) | (h,l^\pm,u)\rangle \ ,
\end{equation}
where $| (h,k^+/2,l^-,u)\rangle$ denotes a highest weight state that is annihilated by all positive modes 
as well as $A^{\ast +}_0$, and the $Q^{--}_{r}$ generators are analogeously defined to (\ref{Gcomp}). 
Its norm equals 
\begin{equation}
|\!|{\cal N}_2 |\!| = 4 \Bigl[ 
h - \frac{k^+ l^- + k^- l^+ + u^2 + (l^+-l^-)^2}{k^++k^-} \Bigr] \ , 
\end{equation}
and thus unitarity requires that we have the `BPS'-bound
\begin{equation}\label{A28}
h \geq \frac{1}{k^++k^-} \, \Bigl[ k^+ l^- + k^- l^+ + u^2 + (l^+-l^-)^2 \Bigr] \ .
\end{equation}
Note that this bound differs from the the corresponding BPS bound of the wedge algebra 
$D(2,1|\alpha)$, see (\ref{A30}); apart from the additional $u^2$ term there is in particular also the
$(l^+-l^-)^2$ term.

\subsection{The Non-linear Algebra $\tilde{A}_\gamma$}\label{app:nonl}

As explained in \cite{Goddard:1988wv}, we can factor out the free fermions and the $\mathfrak{u}(1)$ current
from the large ${\cal N}=4$ algebra $A_\gamma$ to obtain the non-linear $\tilde{A}_\gamma$ algebra. 
The resulting algebra is characterised by the following
commutation relations. First, the levels of the two $\mathfrak{su}(2)$ factors are reduced by $1$, i.e.\ the 
new levels are
\be
\hat{k}^\pm = k^\pm - 1 \ . 
\ee
Thus in terms of the new levels the parameter $\gamma$ is defined as 
\be\label{gammadef}
\gamma \equiv \frac{k^-}{k^+ + k^-} =  \frac{\hat{k}^- + 1}{\hat{k}^+ + \hat{k}^- + 2} \ .
\ee
Similarly, the central charge that appears in the Virasoro algebra is reduced by $3$, so we have 
\be\label{chat}
\hat{c} = \frac{6 k^+ k^-}{k^+ + k^-} - 3 = \frac{6 \hat{k}^+ \hat{k}^- + 3 (\hat{k}^+ + \hat{k}^-)}{\hat{k}^+ + \hat{k}^- +2}   \ . 
\ee
The commutation relations involving the Virasoro and the affine modes are otherwise unmodified, e.g.\ 
(\ref{A.17}) -- (\ref{A.30}) are unchanged, except that the terms proportional to $Q^{\pm\pm}$ are absent. 
However, the structure constants of the supercharge anti-commutation relations get modified; in particular,
$\gamma$ and $(1-\gamma)$ get replaced by 
\be\label{gamma12}
\gamma \mapsto \gamma_1 \equiv \frac{k^--1}{k^+ + k^-}  = \frac{\hat{k}^-}{\hat{k}^+ + \hat{k}^- +2} \ , \quad
(1-\gamma) \mapsto \gamma_2 \equiv \frac{k^+ -1}{k^+ + k^-}  
= \frac{\hat{k}^+}{\hat{k}^+ + \hat{k}^- +2} \ .
\ee
Furthermore, the $c$-parameter that appears in the anti-commutators 
$\{G^{++}_r, G^{--}_s \}$ and $\{G^{+-}_r, G^{-+}_s \}$ is replaced by 
\be
c \mapsto \tilde{c} =  \frac{6 (k^+-1) (k^--1)}{k^+ + k^-}  =  \frac{6 \hat{k}^+ \hat{k}^-}{\hat{k}^+ + \hat{k}^- + 2} 
\ ,
\ee
and non-linear terms (that are bilinear in the currents) appear in all anti-commu\-ta\-tors. For example, the 
first few anti-commutators are 
\be\label{GGpmnon}
\begin{array}{rcl}
{}\{G^{++}_r, G^{++}_s \} & = & - \frac{8}{(\hat{k}^+ + \hat{k}^- +2)} \, (A^{++} A^{-+})_{r+s} \\[4pt]
{}\{G^{++}_r, G^{+-}_s \} & = & 4 (r-s) \,\gamma_1\,  A^{++}_{r+s} 
+ \frac{8}{(\hat{k}^+ + \hat{k}^- +2)} \, (A^{++} A^{-3})_{r+s} \\[4pt]
{}\{G^{++}_r, G^{-+}_s \} & = & 4 (r-s) \, \gamma_2 \, A^{-+}_{r+s} 
+ \frac{8}{(\hat{k}^+ + \hat{k}^- +2)} \, (A^{+3} A^{-+})_{r+s} \\[4pt]
{}\{G^{++}_r, G^{--}_s \} & = & - 4 L_{r+s} - \tfrac{2\tilde{c}}{3} (r^2-\tfrac{1}{4}) \delta_{r,-s} 
- 4 (r-s) \bigl[ \gamma_1 \, A^{+3}_{r+s} + \gamma_2 \, A^{-3}_{r+s} \bigr] \nonumber \\[4pt]
& & +  \frac{4}{(\hat{k}^+ + \hat{k}^- +2)} \, \Bigl( 
(A^{+3} A^{+3}) + \tfrac{1}{2} (A^{++} A^{+-}) + \tfrac{1}{2}(A^{+-} A^{++}) \Bigr)_{r+s} \\[4pt]
& & +  \frac{4}{(\hat{k}^+ + \hat{k}^- +2)} \, \Bigl( 
(A^{-3} A^{-3}) + \tfrac{1}{2} (A^{-+} A^{--}) + \tfrac{1}{2}(A^{--} A^{-+}) \Bigr)_{r+s} \\[4pt]
& & -  \frac{8}{(\hat{k}^+ + \hat{k}^- +2)} \, (A^{+3} A^{-3})_{r+s} \ , 
\end{array}
\ee
etc., where the bilinear currents are normal ordered in the usual manner.
 We should note that the quantum algebra is invariant under the exchange of 
$\hat{k}^+ \leftrightarrow \hat{k}^-$ (or indeed $k^+ \leftrightarrow k^-$), under which 
the two parameters $\gamma_1$ and $\gamma_2$ in (\ref{gamma12}) get interchanged. 
This symmetry corresponds to the classical symmetry $\gamma \leftrightarrow (1-\gamma)$, see
(\ref{gammadef}). 

We should also mention that given $\gamma$ and $c$, say, there are two solutions for 
$(\hat{k}^+,\hat{k}^-)$ for which  $\gamma$ takes the value (\ref{gammadef}), and 
$c=\hat{c}$ in (\ref{chat}).
However, the parameters $\hat{k}^\pm$ appear explicitly
in the commutation relations of the non-linear ${\cal N}=4$ algebra, namely as the levels of the two 
affine $\mathfrak{su}(2)$ algebras. Thus the two corresponding quantum algebras are not
equivalent to one another. Furthermore, since all the structure constants of the non-linear ${\cal N}=4$ algebra 
are determined in terms of $\hat{k}^\pm$,  it follows that the exchange 
of $\hat{k}^+ \leftrightarrow \hat{k}^-$ is the {\em only} triality-like
symmetry of the non-linear ${\cal N}=4$ algebra.

\subsection{The BPS Bound for the Non-linear $\tilde{A}_{\gamma}$ Algebra}

For the case of the non-linear algebra $\tilde{A}_\gamma$, the free fermions that 
appear in (\ref{A26}) are not part of the algebra, and hence the relevant vector is 
\begin{equation}\label{A26n}
{\cal N}_2 =  G^{--}_{-1/2} \, | (h,l^\pm,u)\rangle \ .
\end{equation}
Applying $G^{++}_{1/2}= - (G^{--}_{-1/2})^\dagger$ we obtain the BPS bound 
\be\label{BPSnon}
h \geq \frac{1}{\hat{k}^+ + \hat{k}^- + 2} \, \Bigl[
 (\hat{k}^++1) \, l^- + (\hat{k}^- + 1) \, l^+  + (l^+ - l^-)^2 \Bigr] \ .
\ee
Note that this bound has essentially the same structure as that for the linear $A_\gamma$ algebra,
see eq.~(\ref{A28}), the only difference being the shift of the levels and the absence of the $u^2$ term.

\section{Representations of the Coset Algebra}\label{app:reps}

Let us calculate the conformal dimensions of the coset 
representations $([2,0,\ldots,0];0)$ and $([0,1,0,\ldots,0];0)$ that arise as two-particle states
from $({\rm f};0)$. These states simply appear in the ground state representations of the numerator
algebra. We have the decompositions
\be
[2,0,\ldots,0]^{(N+2)} = ([2,0,\ldots,0], {\bf 1})_{2} \ + ({\bf N},{\bf 2})_{-N/2+1} + ({\bf 1}, {\bf 3})_{-N} 
\ee
and 
\be\
[0,1,\ldots,0]^{(N+2)} = ([0,1,0,\ldots,0], {\bf 1})_{2} \ + ({\bf N},{\bf 2})_{-N/2+1} + ({\bf 1}, {\bf 1})_{-N}  \ .
\ee
The state $([2,0,\ldots,0];0)$ has $l^+=1$, $l^-=0$ and $\hat{u}=-N$, while 
$([0,1,0,\ldots,0];0)$ has $l^+=l^-=0$, $\hat{u}=-N$. From the coset point of view, the conformal weights therefore 
equal
\begin{eqnarray}
h([2,0,\ldots,0];0)  & = &  \frac{C([2,0,\ldots,0])^{\mathfrak{su}(N+2)}}{k+N+2} - 
\frac{N^2}{N(N+2)(N+k+2)}  \notag \\
& = & \frac{N+2}{k+N+2}   \label{4.44} 
\end{eqnarray}
and
\begin{eqnarray}
h([0,1,0,\ldots,0];0)  & = &  \frac{C([0,1,0,\ldots,0])^{\mathfrak{su}(N+2)}}{k+N+2} - 
\frac{N^2}{N(N+2)(N+k+2)}  \notag \\
& = & \frac{N}{k+N+2} \ . \label{4.45}
\end{eqnarray}
On the other hand, the relevant BPS bounds are 
\begin{eqnarray}
h(l^+=1,l^-=0,u=0)_{\rm BPS} & = &  \frac{N+1+1}{N+k+2}  = \frac{N+2}{N+k+2} \label{4.46} \\
h(l^+=0,l^-=0,u=0)_{\rm BPS} & = &  0 \ .  \label{4.47}
\end{eqnarray}
Thus  $([2,0,\ldots,0];0)$ saturates the BPS bound, whereas $([0,1,0,\ldots,0];0)$ does not.
\medskip

The analysis for the representations $(0;[2,0,\ldots,0])$ and $(0;[0,1,0,\ldots,0])$ is similar. These states arise from
the symmetric or antisymmetric combination of the states
\be
\psi^{i,\alpha}_{-1/2} \, \psi^{j,\beta}_{-1/2} |0\rangle \ , 
\ee
respectively. Thus the quantum numbers are $l^+=0$, $l^-=0$, $\hat{u}=N+2$, and 
$l^+=0$, $l^-=1$, $\hat{u}=N+2$, respectively. (Because of the fermionic nature of these oscillators, the roles
of the two representations is reversed.) From the coset point of view, the conformal dimensions equal
\begin{eqnarray}
h(0;[2,0,\ldots,0])  & = & 1 -  \frac{C([2,0,\ldots,0])^{\mathfrak{su}(N)}}{k+N+2} - \frac{(N+2)^2}{N(N+2)(N+k+2)}  \notag \\
& = & \frac{k}{k+N+2}  \label{B8}
\end{eqnarray}
\begin{eqnarray}
h(0;[0,1,0,\ldots,0])  & = &  1 - \frac{C([0,1,0,\ldots,0])^{\mathfrak{su}(N)}}{k+N+2} - \frac{(N+2)^2}{N(N+2)(N+k+2)}  \notag \\
& = & \frac{k+2}{k+N+2} \ .
\end{eqnarray}
On the other hand, the relevant BPS bounds are 
\begin{eqnarray}
h(l^+=0,l^-=1,u=0)_{\rm BPS} & = &  \frac{k+1+1}{N+k+2}  = \frac{k+2}{N+k+2} \label{B10} \\
h(l^+=0,l^-=0,u=0)_{\rm BPS} & = & 0 \ .\label{B11}
\end{eqnarray}
Thus it follows that $(0;[0,1,0,\ldots,0])$ is BPS, while $(0;[2,0,,0,\ldots,0])$ is not.

\subsection{The General BPS States}\label{app:Bgen}

For the general case, the state $([p,0,\ldots,0];0)$ saturates the BPS bound with $l^+=\tfrac{p}{2}$ and
$l^-=0$. Indeed, this state has $\hat{u}= \tfrac{pN}{2}$, and 
\be
C^{(N+2)} \bigl([p,0,\ldots,0]\bigr) = \frac{1}{2} \, p \, (N+1) \, \frac{p+N+2}{N+2} \ , 
\ee
and hence
\begin{eqnarray}
h([p,0,\ldots,0];0) & = & \frac{p(N+1)(p+N+2)}{2(N+2)(k+N+2)} - \frac{p^2 N^2}{4 N (N+2) (N+k+2)} \notag \\
& = & \frac{1}{N+k+2} \Bigl( \frac{p}{2} (N+1) + \frac{p^2}{4} \Bigr) = h_{\rm BPS}(l^+=\tfrac{p}{2},l^-=0,u=0) \ .\notag
\end{eqnarray}

Similarly, the state $(0;[0^{p-1},1,0,\ldots,0])$ saturates the BPS bound with $l^+=0$ and $l^-=\tfrac{p}{2}$. 
It appears in the $p$-fold anti-symmetric product of the free fermion generators and therefore has 
$\hat{u}= \tfrac{p(N+2)}{2}$. Since 
\be
C^{(N)} \bigl( [0^{p-1},1,0,\ldots,0] \bigr) = \frac{1}{2} \, p \, (N-p) \bigl( 1 + \tfrac{1}{N} \bigr) \ , 
\ee
the corresponding conformal dimension equals
\begin{eqnarray}
h(0;[0^{p-1},1,0,\ldots,0]) & = & \frac{p}{2} - \frac{p(N-p)(N+1)}{2N(k+N+2)} - \frac{p^2 (N+2)^2}{4 N (N+2) (N+k+2)} \notag \\
& = & \frac{1}{N+k+2} \Bigl( \frac{p}{2} (k+1) + \frac{p^2}{4} \Bigr) = h_{\rm BPS}(l^+=0,l^-=\tfrac{p}{2},u=0) \ .\notag
\end{eqnarray}

\bibliographystyle{JHEP}

\end{document}